\documentclass{tCPH2e}

\usepackage{epstopdf}
\usepackage{subfigure}

\usepackage{graphics,tikz}
\usetikzlibrary{matrix,arrows,decorations.pathmorphing,decorations.pathreplacing,decorations.markings,backgrounds,positioning,fit,petri,positioning,shapes,snakes,fadings,automata,shadows}
\usepackage[thicklines]{cancel}
\usetikzlibrary{calc,shapes.callouts,shapes.arrows}
\usepackage{amsmath,amssymb,euscript,array,mathrsfs,ctable,marvosym}

\theoremstyle{plain}

\theoremstyle{definition}

\theoremstyle{remark}

\def\SCH{Schr\"odinger\ }

\def\BH{{\cal H}}

\newcommand{\IM}{\operatorname{Im}}
\newcommand{\RE}{\operatorname{Re}}

\def\bra#1{\langle #1|}
\def\ket#1{| #1\rangle}

\def\LCATH{\raisebox{-2pt}{\begin{tikzpicture}[scale=0.14]
\draw (0,0) ellipse (1.8cm and 1.2cm);
\draw [-] (1,0) -- (2.3,0);
\draw [-] (-1,0) -- (-2.3,0);
\draw [-] (1,-0.2) -- (2.2,-0.5);
\draw [-] (-1,-0.2) -- (-2.2,-0.5);
\draw [-] (1,0.2) -- (2.2,0.5);
\draw [-] (-1,0.2) -- (-2.2,0.5);
\filldraw[black] (0.4,0.3) circle (0.1cm);
\filldraw[black] (-0.4,0.3) circle (0.1cm);
\draw[-] (-0.5,-0.6) to[out=-30,in=210] (0.5,-0.6);
\draw[-] (0.8,1.1) -- (1.1,1.8) -- (1.4,0.8);
\draw[-] (-0.8,1.1) -- (-1.1,1.8) -- (-1.4,0.8);
\end{tikzpicture}
}\!}

\def\DCATH{\raisebox{-2pt}{\begin{tikzpicture}[scale=0.14]
\draw (0,0) ellipse (1.8cm and 1.2cm);
\draw [-] (1,0) -- (2.3,-0.5);
\draw [-] (-1,0) -- (-2.3,-0.5);
\draw [-] (1,-0.2) -- (2.2,-1);
\draw [-] (-1,-0.2) -- (-2.2,-1);
\draw [-] (1,0.2) -- (2.2,0);
\draw [-] (-1,0.2) -- (-2.2,0);
\draw[-] (0.2,0.1) -- (0.6,0.5);
\draw[-] (0.2,0.5) -- (0.6,0.1);
\draw[-] (-0.2,0.1) -- (-0.6,0.5);
\draw[-] (-0.2,0.5) -- (-0.6,0.1);
\draw[-]  (-0.5,-0.6) to[out=30,in=150] (0.5,-0.6);
\draw[-] (0.8,1.1) -- (1.1,0.6) -- (1.4,0.8);
\draw[-] (-0.8,1.1) -- (-1.1,0.6) -- (-1.4,0.8);
\end{tikzpicture}
}\!}

\newcommand{\EQ}[1]{\begin{equation}\begin{split} #1
\end{split}\end{equation}}

\begin{document}



\title{\textit{Copenhagen Quantum Mechanics}}

\author{
\name{Timothy J. Hollowood}
\affil{Department of Physics, Swansea University, Swansea SA2~8PP, U.K.}
\received{September 2015}
}

\maketitle

\begin{abstract}
In our quantum mechanics courses, measurement is usually taught in passing, as an ad-hoc procedure involving the ugly collapse of the wave function. No wonder we search for more satisfying alternatives to the Copenhagen interpretation. 
But this overlooks the fact that the approach fits very well with modern measurement theory with its notions of the conditioned state and quantum trajectory. In addition, what we know of as the Copenhagen interpretation is a later 1950's development and some of the earlier pioneers like Bohr did not talk of wave function collapse. In fact, if one takes these earlier ideas and mixes them with later insights of decoherence, a much more satisfying version of Copenhagen quantum mechanics emerges, one for which the collapse of the wave function is seen to be a harmless book keeping device.
Along the way, we explain why chaotic systems lead to wave functions that spread out quickly on macroscopic scales implying that \SCH cat states are the norm rather than curiosities generated in physicists' laboratories. We then describe how the conditioned state of a quantum system depends crucially on how the system is monitored illustrating this with the example of a decaying atom monitored with a time of arrival photon detector, leading to Bohr's quantum jumps. On the other hand, other kinds of detection lead to much smoother behaviour, providing yet another example of complementarity.
Finally we explain how classical behaviour emerges, including classical mechanics but also thermodynamics.\end{abstract}

\begin{keywords}
measurement problem; Copenhagen interpretation; quantum trajectories; decoherence; quantum jumps
\end{keywords}

\section{Introduction}\label{s1}

When we learn quantum mechanics (QM) at first it seems so alien, but then we find that it is a theory of vectors and the central dynamical equation, the \SCH equation, is a first order differential equation (in time at least). Systems evolve smoothly and deterministically in time. Then we learn how powerful it is at explaining microscopic phenomena, like atomic structure. And then we have half a lecture on measurement. More specifically, we are told that when we measure an observable $\hat A$ for the system in the state $\ket{\Psi}$ the outcomes are one of the eigenvalues $a_i$ of $\hat A$ realized with a probability $p_i=|\bra{\psi_i}\Psi\rangle|^2$, where $\ket{\psi_i}$ is the corresponding eigenvector, and after the measurement the system has changed, or collapsed, into the state $\ket{\psi_i}$. The lecturer is a little embarrassed, the \SCH equation has been suspended for a brief moment and quantum system undergoes a special form of stochastic evolution. This is all part of the weirdness of QM, no one really understands it, the lecturer says, and now it's time time to move on to ...

This is pretty much how we learn the Copenhagen interpretation of QM and the reason why we can get away with confining it to half a lecture is that it works so well that we can turn our attention to the microscopic systems themselves. After all, particle physicists routinely apply this simple notion of a quantum measurement to link their theories to what they actually detect at the LHC. The difficult bit is the underlying quantum field theory, not the measurement part. So, although we are generally not satisfied with the standard Copenhagen interpretation, we can put our reservations firmly on the back burner and spend our time thinking about microscopic systems. 

However, technological advances have had an amazing impact on the kinds of measurements that can now be undertaken. Whilst the standard measurement theory was good for describing ensembles of identical measurements, like 
the collision of protons over and over again in the LHC, now, for example, we can make what amount to continuous measurements on single atoms. Systems like this have rich individual trajectories that require something more sophisticated than ensemble averages. These technological advances have driven advances in the theory of measurement itself that build on rather than replace the standard story. What these advances, both in experiment and theory, bring to the fore is the often ignored random face of QM and this begs for a reappraisal of how to incorporate this stochasticity more completely into the formalism.
After all, Geiger counters really do click randomly and photon detectors really do detect a photon at a random time emitted by a single excited atom.

These modern advances, however, do not really quell our unease with the standard interpretation with its
baggage of seemingly subjective observation-induced wave function reduction (collapse) and its dangerous hints of non locality. 
But what we know as the  ``Copenhagen interpretation" is a version of QM that comes to us mainly from Heisenberg in the 1950's, well after the miraculous quantum decade of the 1920's when all the major developments in the theory occurred \cite{Howard1}. Somehow this later version of QM, what became the text-book standard, has become confused with the earlier 1920's QM of the ``Copenhagen School" and its Principal, Niels Bohr. 

This is unfortunate because, this earlier Copenhagen QM (the word interpretation was not used until the 1950's) was different in many respects from the later variant. Bohr, for instance, never talked about the reduction of the wave function and was very careful not to introduce any hint of subjectivity into QM \cite{Howard2}. On the contrary, Bohr, Pauli and others, including the younger Heisenberg, talked about a ``cut" between the measuring device and the microscopic system, a concept that does not seem to feature in discussions of the Copenhagen interpretation today, except loosely in a subjective sense as being between the observer and the observed or between the classical and quantum worlds:
\begin{quote}
To get an observation,  one must therefore cut out a partial system somewhere from the world, and one must make ``statements" or ``observations" just about this partial system. Therefore one destroys the fine connections of phenomena, ...  (Heisenberg 1928 \cite{BohrC}.)
\end{quote}

There are other interesting myths, for example, that our fascination with entanglement is modern and linked to the 
development of quantum information theory. But this is simply not true \cite{Howard3}, entanglement was the hot topic of the 1920's
even though the terminology itself did not get introduced until 1935 by Schr\"odinger \cite{Sch}. For example, in the quote above, Heisenberg uses the phrase ``fine connections of phenomena" in 1928 to describe entanglement. Moreover, entanglement was seen as the principal puzzle of measurement in QM.

The problem with the 1920's approach to measurement of the Copenhagen school is that it was never sufficiently formed or articulated in a way that made an easy ``sell" to later generations, hence our reliance on the later Heisenberg. 
One way to remedy this situation, is to carefully reconstruct the ideas, as Howard has done for Bohr's complementary approach \cite{Howard2}. In this reconstruction, the Copenhagen cut provides a way to understand entangled states in a way that becomes even more compelling when wedded to the modern understanding of decoherence. 
Wave function reduction is then seen to be a harmless book keeping procedure rather than a real physical process: 
\begin{quote}
... it is to be stressed that at first such reductions are not necessary if one includes in the system all means of measurement. But in order to describe observational results theoretically at all, one has to ask what one can say about just a part of the whole system. (Pauli letter to Bohr 1927 \cite{Pauli}.)
\end{quote}

The purpose of this article will be to argue that the more authentic 1920's version of Copenhagen QM has much to recommend it: the nausea associated with wave function collapse can be quelled; observers with their brains and their minds (but not their apparatus) can be returned to neuroscience; and Alice and Bob can finally be retired. The stochastic nature of QM is brought into the foreground and seen to be just as essential to the formalism as the \SCH equation.

Apart from these initial comments, we will not review in any meaningful way the history of QM and the Copenhagen interpretation or give any overview of other so-called ``interpretations". Rather, the goal will be to show that there is a distinct approach here, inspired by the reconstruction of Bohr's complementarity ideas and infused with insights from decoherence. 

Our guiding principles will be that (i) we should take phenomena at face value, so if a photon detector clicks
randomly then there really is something random going on to be described using the best mathematical techniques to hand (i.e.~stochastic processes) and, (ii) observers and their brains, minds or their knowledge, should be no more part of QM than they are of classical mechanics. 

\section{Quantum Measurements}\label{s2}

The characteristic feature of QM is its linearity: states are vectors and so one can add quantum states together to get new states. For example, if we take the two states of a photon, $\ket{h}$ and $\ket{v}$, corresponding to plane polarization in the horizontal and vertical planes (for a photon moving in the other horizontal direction), and make the vector sum
\EQ{
\ket{\Phi}=c_1\ket{h}+c_2\ket{v}\ ,
\label{rnn}
}
we get another state of a single photon with arbitrary polarization, including plane polarized states when $c_1$ and $c_2$ are real, at an angle $\arctan(c_2/c_1)$ to the horizontal axis. For general complex $c_2/c_1$, we get a photon with elliptical polarization.

If quantum mechanics is {\it the\/} universal theory of everything, then nothing should be off limit. In particular, there should  be quantum states corresponding to, say, a cat and, in particular, a live cat $\ket{\LCATH}$ and a dead cat $\ket{\DCATH}$. Both states make perfect macroscopic sense. But now if we make a vector sum as above,
\EQ{
\ket{\Phi}=c_1\ket{\LCATH}+c_2\ket{\DCATH}\ ,
\label{scs}
}
what kind of state do we get? Unlike the photon, where we get a different state of a single photon, now we don't get a partly dead, partly live, cat, rather we have either a live cat or a dead cat with probabilities $|c_1|^2$ and $|c_2|^2$, respectively. How did ``and" turn into ``or" and where do the probabilities come from? The answer will lie in entanglement: ironically the most quantum of phenomena ensures the most classical of phenomena.

\subsection{A simple polarization measurement}\label{s2.0}

The deep problem associated with the \SCH cat states like \eqref{scs} are laid bare in the simplest setting of the measurement of the polarization states of single photons. The set up is shown in the figure below. Photons in some general polarization state, can be written in terms of the horizontally and vertically polarized state as in \eqref{rnn} 
for complex numbers $c_i$ such that $|c_1|^2+|c_2|^2=1$. 
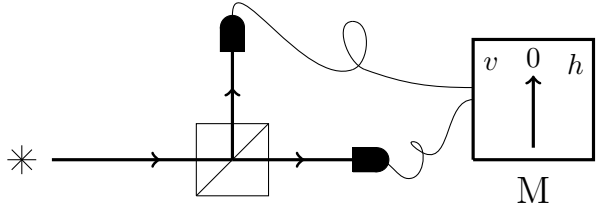
\begin{figure}[ht]
\begin{center}
\begin{tikzpicture}[scale=0.8,decoration={markings,mark=at position 0.6 with {\arrow{>}}}]
\draw[very thick,postaction={decorate}] (-3,0) -- (0,0);
\draw[very thick,postaction={decorate}] (0,0) -- (2,0);
\draw[very thick,postaction={decorate}] (0,0) -- (0,2);
\begin{scope}[xshift=-3.5cm,yshift=0,scale=1.2]
\draw[-] (0,-0.2) -- (0,0.2);
\draw[-] (-0.2,0) -- (0.2,0);
\draw[-] (0.14,-0.14) -- (-0.14,0.14);
\draw[-] (-0.14,-0.14) -- (0.14,0.14);
\end{scope}
\begin{scope}[scale=3,xshift=0cm]
\draw (-0.2,-0.2) -- (-0.2,0.2) -- (0.2,0.2) -- (0.2,-0.2) -- (-0.2,-0.2);
\draw (-0.2,-0.2) -- (0.2,0.2);
\end{scope}
\begin{scope}[scale=1,xshift=2.4cm]
\filldraw[black] (-0.4,-0.2) -- (-0.4,0.2) -- (0,0.2) -- (0,-0.2) -- (-0.4,-0.2);
\filldraw[black] (0,0) circle (0.2cm);
\end{scope}
\begin{scope}[scale=1,xshift=0cm,yshift=2.2cm,rotate=90]
\filldraw[black] (-0.4,-0.2) -- (-0.4,0.2) -- (0,0.2) -- (0,-0.2) -- (-0.4,-0.2);
\filldraw[black] (0,0) circle (0.2cm);
\end{scope}
\begin{scope}[scale=1,xshift=5cm,yshift=1cm]
\draw[very thick] (-1,-1) -- (-1,1) -- (1,1) -- (1,-1) -- (-1,-1);
\draw[very thick,->] (0,-0.8) -- (0,0.4);
\node at (0,0.7) (a1) {$0$};
\node at (-0.7,0.6) (a2) {$v$};
\node at (0.7,0.6) (a3) {$h$};
\node at (0,-1.5) (a4) {\Large M};
\end{scope}
\draw[-] (0,2.4) to[out=90,in=-160] (2,1.5);
\draw[-] (2,1.5) to[out=20,in=0] (2,2.3);
\draw[-] (2,2.3) to[out=-180,in=160] (2.2,1.5);
\draw[-] (2.2,1.5) to[out=-20,in=180] (4,1.2);
\draw[-] (2.4,0) to[out=0,in=-160] (3,-0.3);
\draw[-] (3,-0.3) to[out=20,in=0] (3,0.3);
\draw[-] (3,0.3) to[out=-180,in=160] (3.2,0.1);
\draw[-] (3.2,0.1) to[out=-20,in=180] (4,1);
\end{tikzpicture}
\end{center}
\caption{An apparatus to measure the linear polarization of individual photons using a polarizing beam splitter which reflects horizontally polarized photons upwards and transmits vertically polarized photons.}
\label{f1}
\end{figure}
Photons are incident on a polarizing beam splitter, so that photons in state $\ket{h}$ are reflected and those in state $\ket{v}$ are transmitted. A measuring device $M$ consists of photon detectors and a pointer that indicates the polarization. We take it to have states $\ket{M_0}$, the initial state, and $\ket{M_i}$, $i=1,2$ the final states indicating that a 
photon with polarization $h$ and $v$, respectively, has been detected. It is crucial that the 
states $\ket{M_i}$,  the {\it pointer states\/}, for different $i$ are macroscopically distinguishable. This means that their overlap is vanishingly small $\bra{M_1}M_2\rangle=0$.

$M$ is designed so that when we solve Schr\"odinger's equation for the combined system we find the solutions
\EQ{
&\ket{\Psi(0)}=\ket{M_0}\otimes\ket{h}\longrightarrow\ket{\Psi(T)}=\ket{M_1}\otimes\ket{h}\ ,\\[3pt]
&\ket{\Psi(0)}=\ket{M_0}\otimes\ket{v}\longrightarrow\ket{\Psi(T)}=\ket{M_2}\otimes\ket{v}\ .
}
The linearity of the Schr\"odinger equation then implies that when the polarization is arbitrary
\EQ{
\ket{\Psi(0)}=\ket{M_0}\otimes(c_1\ket{h}+c_2\ket{v})\longrightarrow\ket{\Psi(T)}=c_1\ket{M_1}\otimes\ket{h}+c_2\ket{M_2}\otimes\ket{v}\ ,
\label{gyy}
}
the final state is a superposition of macroscopically distinct states, i.e.~a \SCH cat state. 

Physics is concerned with the phenomenology of real systems and in this case the final state always manifests as either $\ket{M_1}\otimes\ket{h}$ {\it or\/} $\ket{M_2}\otimes\ket{v}$ realized with probabilities given by the Born rule, $|c_1|^2$ 
and $|c_2|^2$, respectively. This can be verified by repeating the experiment many times. In fact, the repetition of experiments a large number of times is often required to verify some underlying quantum prediction. That is why the LHC must allow protons to collide a huge number of times before any meaningful statements can be made regarding a theorist's calculation of the probability that a single collision could produce a Higgs boson or a supersymmetric partner.

The big problem of QM is the tension between the QM state \eqref{gyy} and the observed state:
\begin{center}
\begin{tikzpicture}[scale=1,decoration={markings,mark=at position 0.6 with {\arrow{>}}}]
\node at (-3.5,0) (a1) {$c_1\ket{M_1}\otimes\ket{h}+c_2\ket{M_2}\otimes\ket{v}$};
\node at (3.8,0.6) (a2) {$\ket{M_1}\otimes\ket{h}\qquad p_1=|c_1|^2$};
\node at (3.8,-0.6) (a3) {$\ket{M_2}\otimes\ket{v}\qquad p_2=|c_2|^2$};
\node at (0,0) (z1) {versus};
\draw[decoration={brace,amplitude=0.5em},decorate,very thick] (1.5,-0.8) -- (1.5,0.8);
\end{tikzpicture}
\end{center}
It is impossible to review all the different approaches that have been taken to this problem; however, the
following are three representative ways of viewing what is happening:

\vspace{0.5cm}
{\bf (1)} The final state $\ket{\Psi(T)}$ describes two branches of reality that co-exist equally in the Hilbert space of the system corresponding to $\ket{M_1}\otimes\ket{h}$ {\it and\/} $\ket{M_2}\otimes\ket{v}$. Within each branch it appears, subjectively, as if that branch is realized with the appropriate probability $|c_i|^2$.

{\bf (2)} During a measurement, a special kind of stochastic evolution occurs such that the final state is either $\ket{M_1}\otimes\ket{h}$ {\it or\/} $\ket{M_2}\otimes\ket{v}$ realized with probabilities $|c_1|^2$ and $|c_2|^2$, respectively.

{\bf(3)} The final state is (in a way to be explained in sect.~\ref{s4}) {\it both\/} $\ket{\Psi(T)}$ and one of the $\ket{M_1}\otimes\ket{h}$ or $\ket{M_2}\otimes\ket{v}$ realized with probabilities $|c_1|^2$ and $|c_2|^2$, respectively.

\vspace{0.5cm}
Option (1) above is a caricature of the many worlds interpretation of QM. 
In this approach, there is only the \SCH equation and so no up-front role for $|c_i|^2$ as probabilities. If the experiment were repeated $N$ times, there would be $2^N$ branches and the number of these branches for which $n$ $h$-polarized photons are recorded is $N!/n!(N-n)!$. The fact that this number bears no relation to $N|c_1|^2$ is a real difficulty.
It is claimed  \cite{Wallace} (in a way that is not very convincing to the author) that the probabilities $|c_i|^2$ somehow subjectively emerge for observers embedded in the branches. 

Option (2), a caricature of the standard Copenhagen interpretation, requires some mixture of unitary evolution interspersed with stochastic evolution when measurements are made. The change $\ket{\Psi(T)}$ to one of the possibilities $\ket{M_1}\otimes\ket{h}$ and $\ket{M_2}\otimes\ket{v}$, the famous reduction, or collapse, of the state vector, is assumed to take place during the measurement. As such, the blending of two distinct modes of evolution looks rather ad hoc and invites a subjective element into QM (observer induced reduction). 

More worryingly it seems to imply some kind of action-at-a-distance, a point that can be made by considering a generalization of the measurement with photons in the state, e.g.
\EQ{
c_1\ket{h}_1\otimes\ket{h}_2+c_2\ket{v}_1\otimes\ket{v}_2\ .
}
We now have a photon version of the famous EPR thought experiment \cite{EinsteinPodolskyRosen:1935cqmdprbcc}.
The second photon propagates away from the first so that when the measurement is made on the polarization of the first photon, the second photon is space-like separated from it. If the outcome of the measurement indicates $h$, then the reduction of the state vector involves the instantaneous replacement
\EQ{
c_1\ket{M_1}\otimes\ket{h}_1\otimes\ket{h}_2+c_2\ket{M_2}\otimes\ket{v}_1\otimes\ket{v}_2
\longrightarrow c_1\ket{M_1}\otimes\ket{h}_1\otimes\ket{h}_2\ .
}
It looks like this has changed the state of the second photon. Further investigation, however, shows that the effect on the second photon cannot be used to actually send messages (there is no way to force the second photon into a particular state $\ket{h}$ or $\ket{v}$), so is not directly a breakdown of relativistic causality. Nevertheless the effect seems to involve some kind of ``spooky action-at-a-distance".

The strength of the second option (2) is that it places the stochastic element of QM to the fore, rather than denying it and then trying to sneakily slip it back in later, as the many worlds interpretation apparently must do.
We contend that quantum randomness is an objective feature of the world out there where photon detectors do 
register individual photons at random times and Geiger counters really do click randomly; indeed, quantum systems are the best known sources of random numbers. 

\subsection{Generalized measurement theory}\label{s2.1}

Before we explain what option (3) is in sect.~\ref{s4}, let us first open out our discussion of measurement.
Since the early days of QM, the kinds of measurements that can be undertaken has mushroomed in a way that would have been the stuff of dreams to the pioneers of the theory. 
The simple discussion is inadequate to describe more general measurements with real
apparatus and their unavoidable inefficiencies.

The measuring device must be engineered to have a particular interaction with the microscopic system, so that the solution of the Schr\"odinger equation, for an initial state at $t=0$ of the form $\ket{\Phi(0)}=\ket{M_0}\otimes\ket{\Phi}$, has, at some short time later $T$, after which the measurement could be said to have been completed, the entangled form
\EQ{
\ket{\Psi(T)}=\sum_i\ket{M_i}\otimes\hat \Omega_i\ket{\Phi}\ .
\label{fin}
}
This provides a definition of the {\it measurement operators\/} $\hat \Omega_i$ which 
are properties of the measuring device and describe its effect on the microscopic system when the $i^\text{th}$ outcome 
is registered. The states $\ket{M_i}$ are the pointer states of $M$, macroscopically distinct and so having zero inner product.

Unitary evolution, along with the orthogoniality of the pointer states, requires the completeness relation
\EQ{
1=\bra{\Psi(T)}\Psi(T)\rangle=\sum_{ij}\underbrace{\bra{M_i}M_j\rangle}_{=\,\delta_{ij}}\bra{\Phi}\hat \Omega_i^\dagger\hat \Omega_j\ket{\Phi}
=\bra{\Phi}\sum_i\hat \Omega_i^\dagger\hat \Omega_i\ket{\Phi}
\quad\implies \sum_i\hat\Omega_i^\dagger\hat\Omega_i=1\ ,
\label{com}
}
since the initial microscopic state is arbitrary.
This means mathematically that the set of operators $\{\hat\Omega_i^\dagger\hat\Omega_i\}$ form a POVM: 
\begin{quote}{\bf POVM:} For our purposes, a POVM (positive, operator-valued measure) is simply a set of operators $\{\hat\Pi_i\}$ that is complete, in the sense that $\sum_i\hat\Pi_i=1$, and for which the expectation values $\bra{\Phi}\hat\Pi_i\ket{\Phi}$ are positive real numbers $p_i$. Therefore, the $p_i$ can be given the interpretation as probabilities since $p_i\geq0$ and $\sum_ip_i=1$. But note that, in general, there is no requirement that the operators $\hat\Pi_i$ either commute or are projection operators.
\end{quote}
\begin{figure}[ht]
\begin{center}
\begin{tikzpicture}[scale=1,decoration={markings,mark=at position 0.6 with {\arrow{>}}}]
\node at (-2.4,0) (a1) {$\ket{M_0}\otimes\ket{\Phi}$};
\node at (4.2,0) (a3) {$\ket{M_i}\otimes\hat\Omega_i\ket{\Phi}\qquad p_i=\bra{\Phi}\hat\Omega_i^\dagger\hat\Omega_i\ket{\Phi}$};
\draw[very thick,postaction={decorate}] (-1,0) -- (0,0);
\draw[very thick,postaction={decorate}] (0,0) -- (1,0);
\draw[very thick,postaction={decorate}] (0,0) -- (1,1);
\draw[very thick,postaction={decorate}] (0,0) -- (1,0.5);
\draw[very thick,postaction={decorate}] (0,0) -- (1,-0.5);
\draw[very thick,postaction={decorate}] (0,0) -- (1,-1);
\end{tikzpicture}
\end{center}
\caption{A generalized measurement is described by a set of measuring operators $\hat\Omega_i$ that describe the action of the measuring device on the microscopic system when the $i^\text{th}$ outcome is registered.}
\label{f2}
\end{figure}
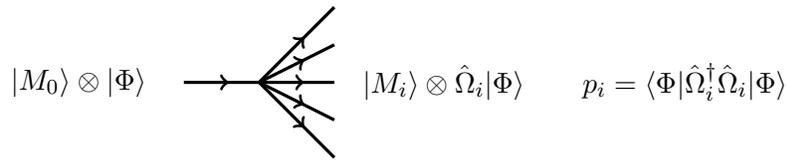

If quantum mechanics is to give rise to classical mechanics in the macroscopic limit, i.e.~give a faithful, phenomenological description of the world, then, clearly, it is necessary to add some additional rules to the theory. In the conventional Copenhagen interpretation, this is what the reduction of the state vector achieves. 
At the end of the measurement, only one of the components $\ket{M_i}\otimes\hat\Omega_i\ket{\Phi}$ is actually realized in a completely random way with probability
\EQ{
p_i=\bra{\Phi}\hat\Omega_i^\dagger\hat\Omega_i\ket{\Phi}\ .
\label{ppr}
}
The completeness condition \eqref{com} ensures that the probabilities add up to unity $\sum_ip_i=1$. 

\subsection{Quantum trajectories}\label{s2.2}

Usually we are not interested in the measuring device and only care about the microscopic system. In that case, if we remove the measuring device from the discussion, the evolution of the microscopic system becomes {\it conditioned\/} by the result of the measurement which means that the state must jump
\EQ{
\ket{\Phi}\longrightarrow\hat\Omega_i\ket{\Phi}\ .
\label{col}
} 
Of course this is the Copenhagen interpretation's reduction of the system's state vector.
This description of the conditioned state lies behind the modern theory of measurement and control theory in QM and provides a very successful description of systems in the lab \cite{WMbook,PK}. 

The situation becomes particularly rich and interesting when it is used to describe a sequence of measurements. The time evolution of the state of the  microscopic system conditioned on the results of such a sequence of measurements is called a {\it quantum trajectory}. If the measurements are discrete events, then this consists of smooth unitary evolution  interspersed with stochastic jumps at the times of the measurements:
\begin{figure}[ht]
\begin{center}
\begin{tikzpicture}[scale=0.6,decoration={markings,mark=at position 0.6 with {\arrow{>}}}]
\draw[very thick,postaction={decorate}] (-7,1) -- (-6,1);
\draw[very thick,postaction={decorate}] (-6,1) -- (-2,-1);
\draw[very thick,postaction={decorate}] (-2,-1) -- (1,0.5);
\draw[very thick,postaction={decorate}] (1,0.5) -- (5,-1.3);
\draw[very thick,postaction={decorate}] (5,-1.3) -- (6,-0.3);
\node at (-6,-3.5) (s1) {$\hat\Omega_{i_1}\ket{\Phi}$};
\node at (-1.5,-3.5) (s2) {$\hat\Omega_{i_2}U_{21}\hat\Omega_{i_1}\ket{\Phi}$};
\node at (4,-3.5) (s3) {$\hat\Omega_{i_3}U_{32}\hat\Omega_{i_2}U_{21}\hat\Omega_{i_1}\ket{\Phi}$};
\draw[->] (s1) -- (-5.8,-2.5);
\draw[->] (s2) -- (-1.8,-2.5);
\draw[->] (s3) -- (1.3,-2.5);
\draw[densely dashed] (-6,2) -- (-6,-1.8);
\draw[densely dashed] (-2,2) -- (-2,-1.8);
\draw[densely dashed] (1,2) -- (1,-1.8);
\draw[densely dashed] (5,2) -- (5,-1.8);
\node at (-6,-2.1) {$t_1$};
\node at (-2,-2.1) {$t_2$};
\node at (1,-2.1) {$t_3$};
\node at (5,-2.1) {$t_4$};
\begin{scope}[scale=1,xshift=-6cm,yshift=1cm]
\draw[postaction={decorate}] (-1,0) -- (0,0);
\draw[postaction={decorate}] (0,0) -- (1,0);
\draw[postaction={decorate}] (0,0) -- (1,1);
\draw[postaction={decorate}] (0,0) -- (1,0.5);
\draw[postaction={decorate}] (0,0) -- (1,-1);
\end{scope}
\begin{scope}[scale=1,xshift=-2cm,yshift=-1cm]
\draw[postaction={decorate}] (0,0) -- (1,0);
\draw[postaction={decorate}] (0,0) -- (1,1);
\draw[postaction={decorate}] (0,0) -- (1,-0.5);
\draw[postaction={decorate}] (0,0) -- (1,-1);
\end{scope}
\begin{scope}[scale=1,xshift=1cm,yshift=0.5cm]
\draw[postaction={decorate}] (0,0) -- (1,0);
\draw[postaction={decorate}] (0,0) -- (1,1);
\draw[postaction={decorate}] (0,0) -- (1,0.5);
\draw[postaction={decorate}] (0,0) -- (1,-1);
\end{scope}
\begin{scope}[scale=1,xshift=5cm,yshift=-1.3cm]
\draw[postaction={decorate}] (0,0) -- (1,0);
\draw[postaction={decorate}] (0,0) -- (1,0.5);
\draw[postaction={decorate}] (0,0) -- (1,-0.5);
\draw[postaction={decorate}] (0,0) -- (1,-1);
\end{scope}
\end{tikzpicture}
\end{center}
\caption{The state of the microscopic system conditioned by the results of a sequence of measurements defines a quantum trajectory. The evolution is unitary between the measurements.}
\label{f3}
\end{figure}
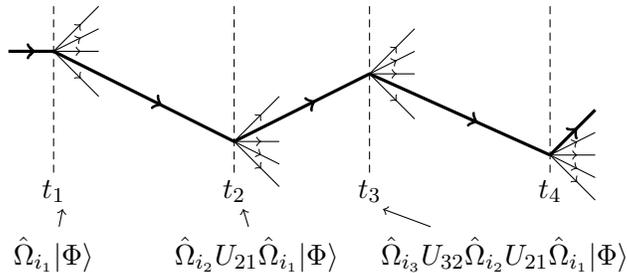
The important point here is that the conditioned state at a certain time is labelled by the trajectory of 
outcomes of all previous measurements  
\EQ{
\ket{\Phi_{i_1,i_2,\ldots,i_n}}=\hat\Omega_{i_n}U_{n,n-1}\hat\Omega_{i_{n-1}}\cdots\hat\Omega_{i_2}U_{21}\hat\Omega_{i_1}\ket{\Phi}\ ,
}
where $U_{21}=U(t_2,t_1)$ is the unitarity evolution operator between $t=t_1$ and $t_2$, etc. The probability of such a trajectory is then
\EQ{
p_{i_1,\ldots,i_n}=\bra{\Phi_{i_1,\ldots,i_n}}\Phi_{i_1,\ldots,i_n}\rangle\ .
}
Some examples of quantum trajectories will be shown in sect.~\ref{s7}.

If one is careful, it is possible to take a limit to model a system being subjected to continuous measurement, the resulting conditioned state then satisfies a stochastic version of the \SCH equation (see \cite{JacobsSteck} for an introduction). This formalism will not be developed here, but it should be emphasized that this stochastic theory of the conditioned state as provided by the Copenhagen interpretation works remarkably well giving a phenomenologically successful description of a large range of real systems in the lab. 

If the above description of measurement was all that the original Copenhagen QM could offer then it is understandable that many would say that it is ad hoc and inadequate. But as we discussed in the introduction, the
Copenhagen interpretation is mainly Heisenberg's 1950's re-working of QM of the Copenhagen school \cite{Howard1}. The original 1920's approach is rather different; in particular, Bohr never talked about the reduction of the wave function and both Bohr and Heisenberg talked about a cut (sometimes called the Heisenberg cut) to be placed between a measuring device and the microscopic system.  

\section{Chaos and the Breakdown of Correspondence Principle}\label{s3}

The central problem in QM is caused by \SCH cat states, macroscopic superpositions. But how common are these states in the real world? Are they just freaks that emerge from a physicist's lab or are they an everyday blight on the classical world? This question is closely related to the 
{\it correspondence principle\/} which describes how classical behaviour can emerge from QM in the limit when typical actions in a problem are $\gg\hbar$ (requiring large quantum numbers).

The usual text-book argument for the correspondence principle starts by recognizing that there are states in the Hilbert space of macroscopic systems that are {\it quasi classical\/} in that they are approximately localized in phase space (position and momentum) up to the fundamental limit set by the uncertainty principle.

For example, consider a particle moving in one dimension. A Gaussian wave packet 
\EQ{
\psi(x)=C\exp\big[-(x-X)^2/4b^2+iPx\big]\ ,
\label{bbs}
}
is spread out in real space over a scale $b$ and in momentum space by $\hbar/2b$. It is, therefore, quasi-classical if $b$ and $\hbar/b$ are small compared with macroscopic scales.
Ehrenfest's theorem then implies that classical equations of motion on phase phase are satisfied at the level of expectation values. For the particle moving in a potential, one has
\EQ{
\frac d{dt}\langle\hat x\rangle=\frac1m\langle\hat p\rangle\ ,\qquad\frac{d}{dt}\langle\hat p\rangle = -\big\langle \frac{dV(\hat x)}{dx}\big\rangle\ ,
}
where $\langle\cdot\rangle$ denotes an expectation value with the wave function $\psi(x,t)$.
As long as the wave packet is very narrow compared with the scale over which the potential varies, then, taking $X(t)$ to be the peak of $|\psi(x,t)|^2$, we have
\EQ{
\langle\hat p\rangle\approx m\frac{dX}{dt}\ ,\qquad\big\langle \frac{dV(\hat x)}{dx}\big\rangle\approx\frac{dV(X)}{dX}\qquad\implies\quad m\frac{d^2X}{dt^2}\approx-\frac{dV}{dX}\ ,
}
implying that the peak of the wave function follows the classical trajectory.

This picture is maintained as long as a state remains narrow compared with the scale over which the potential varies. But even in a free theory,  a quasi-classical state like \eqref{bbs} spreads out in configuration space over time:
\EQ{
\Delta x(t)^2=b^2+\Big(\frac{\hbar t}{2mb}\Big)^2\ .
\label{ytt}
}
For a macroscopic system this spreading is actually slow compared with the age of the universe and may be neglected. Even so, this kind of reasoning seems a poor way to explain the ``classicality" of the world around us. For instance, it would require that the wave function at the big bang was quasi classical to begin with. The issues of why the quantum universe appears classical is the subject of a review by Halliwell \cite{Hal}. Here, we take a different view, pointing out that existence of chaos makes the issue of classicality much more perplexing.

\subsection{Hyperion}\label{s3.1}

However, the reasoning above is a chimera because all real systems are sufficiently complicated to be classically chaotic and the wave functions of such systems spread out much more quickly than \eqref{ytt} leading to a breakdown of Ehrenfest's theorem on remarkably short time scales.

A fascinating example of a chaotic system, much discussed in the quantum chaos literature \cite{Zurek1,Zurek2,Berry,Sch1}, is the intrinsic rotational motion of Hyperion one of the moons of Saturn.
It is an irregular potato-shaped chunk of rock of size $\sim140\ \text{km}$. What picks out Hyperion as particularly special and interesting is that it is not tidally locked to Saturn and its intrinsic rotational motion is chaotic---it tumbles in a way that is unpredictable---as a result of the tidal forces of Saturn and its other moons.

One way to characterize the chaotic behaviour is to consider a family of classical trajectories that start from points in a small compact area in phase space. Area in phase space is, by Liouville's Theorem, preserved under time evolution, but in a chaotic system the perimeter of the area grows exponentially $L(t)=L(0)e^{\lambda t}$. Here, $\lambda$ is a Lyapunov exponent that defines a chaos time scale $t_c=1/\lambda$. As time evolves, the compact region is expanded in some directions and squeezed in other directions and eventually becomes very convoluted with fine tendrils that cover phase space in a complicated way.

For Hyperion, the chaos time scale is $t_c\sim100$ days compared with its intrinsic rotational period of a few days. So over time scales of this order, the motion effectively becomes unpredictable because of the practical uncertainties of specifying the initial state.

\begin{figure}[ht]
\begin{center}
\begin{tikzpicture}[scale=0.9,fill=black!20]
\begin{scope}[xshift=-1.7cm,yshift=-1.5cm]
\begin{scope}[scale=0.6,yshift=-0.2cm]
\shade[ball color=black!25] (0,1) to[out=-10,in=120] (0.4,0.5) to[out=-60,in=50] (0.5,-0.2) to[out=-130,in=20]  (0,-0.9) to[out=-160,in=-80]  (-0.3,-0.4) to[out=100,in=-150]  (-0.4,0.7) to[out=30,in=170] (0,1);
\end{scope}
\end{scope}
\draw (0,0) ellipse (3cm and 2cm);
\draw[->] (-3.8,0) -- (3.5,0);
\draw(1.1,0) -- (-2.6,-2.15);
\draw[very thick,->] (-1.7,-1.6) -- (-1.7,0.7);
\draw[very thick,->] (-1.7,-1.6) -- (-1.15,-2.4);
\draw[very thick,dotted] (-1.7,-1.6) -- (-2.8,0);
\draw (0,0) arc (-180:-149:1.1);
\draw (-3.3,0) arc (-180:-57:0.5);
\node at (-1.7,1) {$I_3$};
\node at (-1,-2.6) {$I_1$};
\node at (-1.1,0.6) {$\ell$};
\node at (-0.4,-0.3) {$\theta(t)$};
\node at (-0.3,-1.2) {$r(t)$};
\node at (-3.4,-0.6) {$\phi$};
\draw[fill] (1.1,0) circle (0.5cm); 
\draw[very thick] (1.1,0) ellipse (1cm and 0.35cm);
\draw[very thick] (1.1,0) ellipse (0.9cm and 0.29cm);
\draw[very thick] (1.1,0) ellipse (0.7cm and 0.2cm);
\draw[fill] (0.6,0) arc (180:0:0.5);
\begin{scope}[yscale=0.5]
\draw[very thick,->] (-1.82,0.8) arc (-250:70:0.3);
\end{scope}
\end{tikzpicture}
\end{center}
\caption{The simplified model of the spin orbit dynamics of Hyperion. The elliptical orbit of the moon around the planet is described by the angle $\theta(t)$ and radius $r(t)$ which are fixed  functions. The satellite rotates with an angular momentum $\ell$ lying perpendicular to the orbital plane in a direction defined by $I_3$, the maximal principal momentum of inertia, $I_3>I_2>I_1$. The rotation is parametrized by the angle $\phi$ between the semi-major axis of the orbit ellipse and the axis defined by the smallest moment of inertia $I_1$.}
\label{f4}
\end{figure}
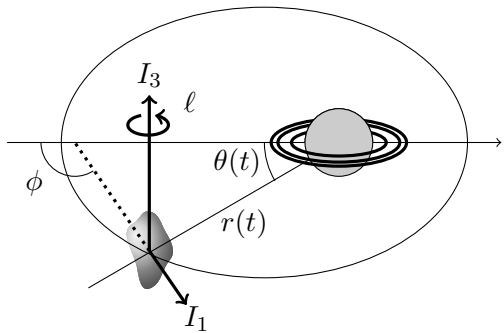

A simplified model of Hyperion, from Wisdom et al \cite{WPM}, is illustrated in fig.~\ref{f4}. One focusses on the spin-orbit coupling between the intrinsic rotational motion of the moon along an axis perpendicular to the plane of its elliptical orbit around Saturn, and the orbital motion itself. The orbit is fixed and specified by the functions $r(t)$ and $\theta(t)$. The problem is then effectively one dimensional with a phase space parametrized by the angle $\phi$ and the angular momentum $\ell$.

The Hamiltonian of the model takes the form
\EQ{
H=\frac{\ell^2}{2I_3}-\frac{3\pi^2}{T^2}\left(\frac a{r(t)}\right)^3(I_2-I_1)\cos(2(\phi-\theta(t)))\ ,
\label{hham}
}
where $T$ is the orbital period and $a$ is the semi-major axis of the orbit and $I_i$ are principal moments of inertial with $I_3>I_2>I_1$.

As a quantum system, with commutator $[\hat\phi,\hat\ell]=i\hbar$, the state vector for the rotational motion can be written in terms of the angular momentum eigenstates:
\EQ{
\ket{\Psi}=\sum_{m=-\infty}^\infty c_m\ket{m}\ ,\qquad \hat\ell\ket{m}=m\hbar\ket{m}\ .
}
It is convenient to introduce position and momentum coordinates $x=R\phi$ and $p=\ell/R$, where $R$ is the moon's
intrinsic average radius.

In the Hilbert space, there are a set of special {\it coherent\/} states 
\EQ{
\ket{x,p}={\cal N}\sum_{m=-\infty}^\infty e^{-(\delta x/\hbar)^2(p -m\hbar/R)^2}e^{imx/R}\ket{m}\ ,
}
which have minimal spread $\delta x\,\delta p=\hbar/2$. The scale $\delta x$ is fixed but arbitrary. The states are (over-)complete in the sense that
\EQ{
\int \frac{dx\,dp}{2\pi\hbar}\,\ket{x,p}\bra{x,p}=1\ .
\label{zza}
}
These states are quasi classical when $\delta x$ is much smaller than the scale over which the potential varies, which in the present problem is order $R$ the radius.

Any state in the Hilbert space can be though of as a density in phase space defined by
\EQ{
\rho(x,p)=\big|\bra{x,p}\Psi\rangle\big|^2\ .
}
This is a bona-fide, i.e.~positive, normalized density on phase space known as the Husimi function \cite{Husimi} of the pure state $\ket{\Psi}$. (Interestingly, quantum mechanics can be completely reformulated in terms of phase space densities, usually the more popular related Wigner function: for a nice historical introduction and original papers see \cite{ZFC}.)

If we start the system off in a quasi-classical state $\ket{x,p}$, we can expect Ehrenfest's theorem to apply and the evolution to follow the classical motion for a while, i.e.~the quantum density $\rho(x,p)$ will follow the classical density on phase space. Due to the exponential spreading of the area's perimeter in the classical system, the corresponding quantum density will likewise spread out. Ehrenfest's theorem will obviously break down when the spreading becomes of the order $R$. We can get an estimate of the spreading time, following \cite{Zurek2}, by assuming that the initial quasi-classical state has an uncertainty of its momentum given by the scale of thermal fluctuations of Hyperion. The surface temperature is roughly $T\sim100\ \text{K}$, mass $m\sim10^{19}\ \text{kg}$ and $\delta p\sim\sqrt{mk_BT}$; hence $\delta x=\hbar/\sqrt{mk_BT}$. In the classical system, a region in phase space is expanded  in the spatial direction and squeezed in the momentum direction, at least initially. Making an approximation that this exponential spreading in the position direction continues until it reaches the spatial size of the phase space, we have
$R/\delta x=\exp[\lambda t_q]$. Hence, given $\lambda=1/t_c$, we find that $t_q$ depends only logarithmically on $\hbar$
\EQ{
t_q\thicksim t_c\log\Big(\frac{R\sqrt{mk_BT}}\hbar\Big)\ .
}
This kind of logarithmic dependence on $\hbar$ is the characteristic feature of the much more careful analyses of \cite{BZ,BB,Zurek2}.

For Hyperion, this yields an astonishing
\[
t_q\thicksim 24\ \text{years}!
\]
This timescale for a quasi-classical state to becomes smeared out over phase space and Ehrenfest's theorem to break down is remarkably small because of the logarithmic dependence on $\hbar$.

So if we thought that the macroscopic superpositions produced by quantum measurements, Schr\"odinger cat states, were somehow very rare beasts, it is sobering to realize that any chaotic system---essentially any system in the real world---will be in a macroscopic superposition. This includes not just Hyperion's rotation, but also the orbital motion of all the planets in the solar system. So the solar system should be in a highly quantum state.

\section{The Copenhagen Cut}\label{s4}

The presentation of a quantum measurement we have described is really based on Heisenberg's later
1950's version of the Copenhagen interpretation with its focus on observation-induced wave function collapse. However, this does not do justice to the original 1920's QM and, in particular, to Bohr who never talked about the collapse of the wave function and never conceded a subjective element in QM. Bohr, Pauli and others, including the younger Heisenberg, did talk about the need for a {\it cut\/} to be placed between the microscopic system and the measuring device. Without this cut, QM could not make any meaningful predictions about the outcome of a measurement. 

The question is how does the cut allow us to make predictions in practice: in short how does the cut work? It is a non-trivial task to tease out the mechanism; however, a careful study of Bohr's original papers by Howard \cite{Howard2}, provides a rather convincing reconstruction of Bohr's ideas. In the end, though, we have to consider the merits of the reconstruction on its own terms. Interestingly, the same idea has also been proposed by others over the years, e.g.~Krips \cite{Krips:1969tpqm}, 
then by many others in the context of modal interpretations of QM. (These are described, for example, in the books \cite{vanFraassen:1991qmmv,Vermaas:1999puqm}. The fact that Bohr's complementarity approach is a type of modal interpretation was noted by Bub and Clifton \cite{BC}). 

It is important also to bear in mind that Bohr and Heisenberg differed about how to think about the cut (discussed in \cite{SC1,SC2}). In particular, Bohr never saw the cut, as has often been claimed, as a division between the microscopic realm described by QM and the macroscopic realm described by classical mechanics. The cut, however, will be seen to give rise to emergent (quasi-)classical properties on one side.

The central problem that the cut allows one to overcome, is how to make sense of a non-separable, i.e.~entangled, state like \eqref{gyy} or \eqref{fin}.  An entangled state like \eqref{gyy} has no classical analogue because it lacks {\it separability\/} and cannot assign separate properties to $M$ or the photon. So if we want to be serious in modelling the real behaviour of systems, somehow we have to extract separable states from the underlying entangled state. This is what the cut achieves.

In our simple model, the cut is placed between the measuring device and the photon. For the moment, we won't ask whether this is natural, but there is no other choice in this simple description.
It is then a fact that the expectation value of {\it any\/} operator $\hat A$ acting {\it only\/} on one side of the cut, i.e.~either solely on $M$, or solely on the photon, can be captured by a ``classical" ensemble average
\EQ{
&\bra{\Psi(T)}\hat A\ket{\Psi(T)}=\bra{\Psi_1}\hat A\ket{\Psi_1}+\bra{\Psi_2}\hat A\ket{\Psi_2}\ ,\\[3pt]
&\text{where}\qquad\ket{\Psi(T)}=\ket{\Psi_1}+\ket{\Psi_2}\ ,\\[3pt]
&\ket{\Psi_1}=c_1\ket{M_1}\otimes\ket{h}\ ,\qquad\ket{\Psi_2}=c_2\ket{M_2}\otimes\ket{v}\ .
\label{kds}
}
Here the ensemble consists of the states $\ket{\Psi_i}$  realized with probability
\EQ{
p_i=\bra{\Psi(T)}\Psi_i\rangle=\bra{\Psi_i}\Psi_i\rangle=|c_i|^2\ .
}
The states $\ket{\Psi_i}$ are completely fixed as long as $|c_1|^2\neq|c_2|^2$. The potential degeneracy problem will
be cured eventually in the more general approach we develop.
Ensembles of the form above are also called {\it mixed states\/} and can also be described using density operators but we will avoid using these in this article.

The crucial point is that the states of the ensemble are separable and so can assign an unambiguous state to both $M$ and the photon. In addition, the states  of $M$ have a well-defined classical interpretation: the photon, say, has $h$ polarization and the apparatus indicated an $h$ photon. 
So picking out the states of the ensemble corresponds to what Heisenberg alludes to in the quote reproduced in sect.~\ref{s1} as the procedure where ``one destroys the fine connections of phenomena", i.e.~converts the entanglement to a separable state.

\subsection{The global and local states}\label{s4.2}

Now comes the crucial part of the argument where we finally explain choice (3) in sect.~\ref{s2}. If real systems only come in the states $\ket{M_1}\otimes\ket{h}$ or $\ket{M_2}\otimes\ket{v}$ after measurement, then one way to make sense of this is to assert that there are two useful notions of  ``state'': the original ``global state" $\ket{\Psi(T)}$ along with the ``local state", describing the system associated to the {\it context\/} provided the cut. The latter are the states of the classical ensemble that give the same expectation values for observables acting only on one side of the cut, that is $\ket{M_1}\otimes\ket{h}$ or $\ket{M_2}\otimes\ket{v}$, realized with the associated probabilities $|c_i|^2$. 

The idea is simple: the cut picks out two complementary subsystems, here $M$ and the photon, and with respect to that choice there is another notion of state. It is the local state that describes the actual state of the subsystems on either side of the cut, so $M$ and the photon. In this way, the state of the photon becomes correlated with the state of $M$ in a separable way. Said another way, the state of the photon becomes {\it conditioned\/} on the measurement outcome.

Note the global state is needed because it determines the possible local states, so we cannot remove it from the formalism. However, we will see that sometimes we can change the global state (if we wish) as time evolves without changing future predictions. This is just wave function reduction exposed as a harmless book keeping exercise.

The notion of the local state allows us to effortlessly introduce randomness into QM: if only one local state is realized at any given time then it is meaningful to talk about the probability for that particular local state being realized. Having two notions of state, or two {\it modalities\/}, is also a central feature of the class of {\it modal\/} interpretations of QM \cite{vanFraassen:1991qmmv,Vermaas:1999puqm}. In that context, the terminology is global/local=dynamical/value, although other words are also used. We should emphasize that modal QM is a collective name for a whole range of inequivalent ``interpretations" of QM.

One important feature of this proposal is that local states of the microscopic system depend on the {\it context\/}, that is exactly how the measuring device has been designed and what it measures. 
This is Bohr's complementarity in practice.
If we design $M$ to measure circular polarization, for instance, then the local states of the photon will be circularly polarized.  

The cut might not seem natural; however, we believe that it becomes so if the resulting macroscopic description is robust under changing the cut, of which we will have more to say later.

\subsection{Preferred observables}

Another, and ultimately more general, way to think about the cut is that it picks out (or is defined by) a preferred set of observables. On the $M$ side, expectation values of a non-local observable like
\EQ{
\hat A_\text{non-local}=e^{i\alpha}\ket{M_1}\bra{M_2}+e^{-i\alpha}\ket{M_2}\bra{M_1}\ ,
}
for arbitrary phase angle $\alpha$, vanish because of the entanglement with the photon:
\EQ{
\bra{\Psi}\hat A_\text{non-local}\ket{\Psi}=\RE\Big[e^{i\alpha}c_1^*c_2\bra{h}v\rangle\Big]=0\ .
\label{rrf}
}
It is only the local observables that are linear combinations of the projection operators $\ket{M_1}\bra{M_1}$ and $\ket{M_2}\bra{M_2}$ that have non-vanishing expectation values, e.g.
\EQ{
\bra{\Psi}\Big\{\alpha\ket{M_1}\bra{M_1}+\beta\ket{M_2}\bra{M_2}\Big\}\ket{\Psi}=\alpha|c_1|^2\bra{h}h\rangle+\beta|c_2|^2\bra{v}v\rangle=\alpha|c_1|^2+\beta|c_2|^2\ .
}
These projection operators become a preferred set of observables that just happen to have a good classical interpretation.

The preferred observables form a complete set:
\EQ{
\ket{M_1}\bra{M_1}+\ket{M_2}\bra{M_2}=1
} 
and their existence is a universal aspect of entanglement. 
In more complicated situations, we will relax the requirement of strict commutativity and identify the preferred observables with a set of local operators $\{\hat\Pi_i\}$ that approximately commute. However, it is important that the set is complete in the sense that $\sum_i\hat\Pi_i=1$, in other words, they form a {\it quasi-classical\/}, i.e~almost commuting, projective POVM.

The preferred observables in the form of the quasi-classical POVM then determine the local states as
$\ket{\Psi_i}=\hat\Pi_i\ket{\Psi}$. In the present example,
\EQ{
\ket{\Psi_1}=\ket{M_1}\bra{M_1}\Psi\rangle=c_1\ket{M_1}\otimes\ket{h}\ ,\qquad
\ket{\Psi_2}=\ket{M_2}\bra{M_2}\Psi\rangle=c_2\ket{M_2}\otimes\ket{v}\ .
}
The fact that the operators $\hat\Pi_1$ and $\hat\Pi_2$ almost commute follows from the fact that the overlap between the macroscopically distinct states $\bra{M_1}M_2\rangle$ will be exceedingly tiny.

\subsection{A potential problem}\label{s4.4}

The fact that the expectation values of observables on $M$ (or the photon) are captured by an ensemble which involves states that are macroscopically distinct depends crucially on the fact that the states of the photon $\ket{h}$ and $\ket{v}$ in the final state \eqref{gyy} are orthogonal. It is this orthogonality that acts to decohere the state of $M$, i.e.~destroys the non-classical interferences between the states $\ket{M_1}$ and $\ket{M_2}$. 

Unfortunately, this nice observation fails when we consider a generalized measurement described in sect.~\ref{s2.1}.
The problem is that there is no requirement that the final states of the microscopic system $\hat \Omega_i\ket{\Phi}$ are orthogonal and in general the photon does not decohere the state of $M$ in an appropriate quasi-classical basis, i.e.~contrary to \eqref{rrf}, we have
\EQ{
\bra{\Psi}\hat A_\text{non-local}\ket{\Psi}=\RE\Big[c_1^*c_2e^{i\alpha}\bra{\Phi}\hat\Omega^\dagger_1\hat\Omega_2\ket{\Phi}\Big]\neq0\ .
}

As an example, suppose that the measurement operators are
\EQ{
\hat\Omega_1=\sqrt{1-\epsilon^2}\ket{h}\bra{h}+\epsilon\ket{v}\bra{v}\ ,\qquad
\hat\Omega_2=\epsilon\ket{h}\bra{h}+\sqrt{1-\epsilon^2}\ket{v}\bra{v}\ ,
\label{rrw}
}
describing a situation where the polarizing beam splitter is not 100\% efficient but makes errors 
at a level set by $\epsilon$, i.e.~an $h$-photon can be transmitted and registered as $v$, and vice-versa. These operators still commute and are a POVM but they are no longer projection operators. In this case, to linear order in $\epsilon$, the local states of the measuring device are now
\EQ{
\ket{M_1}+\frac{\epsilon}{|c_1|^2-|c_2|^2}\ket{M_2} +\cdots\quad\text{and}\quad \ket{M_2}+\frac{\epsilon}{|c_2|^2-|c_1|^2}\ket{M_1}+\cdots
}
and so are superpositions of macroscopically distinct states: the ensemble is no longer ``classical". 
Fortunately, we have overlooked a very important effect in our simple model as will emerge in the next section.

\section{Decoherence}\label{s5}

Decoherence in a dynamical property of systems in the real world with all their inherent complexity that has important and favourable implications for the reconstruction of Bohr's theory of the cut \cite{SC1,SC2}. The literature on the subject is large but for a limited list see \cite{Zurek:1982ii,Caldeira:1982iu,Joos:1984uk,Unruh:1989dd,Zurek91,PHZ,ZHP,Zurek93,ZP,HSZ,Zurek98b,JZKGKS,Schlosshauer:2003zy}. The central idea is that every realistic macroscopic system cannot be considered in isolation and is always interacting with its environment. Of course, this is the case in classical physics for objects that are in thermal equilibrium. In the quantum case, the novel element is that the interaction with the environment builds up entanglement.

One misconception is that the environment consists of spatially disjoint systems. Drawing a macroscopic boundary around an object is really just a matter of convenience and not of fundamental importance.
Rather the split between a system and the environment is a dividing line based on scale: the macroscopic, or infra red, versus the microscopic, or ultra violet. So a macroscopic object carries its own environment consisting of all its microscopic degrees-of-freedom, 
in addition to external microscopic influences. Other external macroscopic degrees-of-freedom are not really to be included in the environment, we may simply include these objects, if relevant, along with the original object.

In this context, the Copenhagen cut is precisely a way to delineate the macroscopic, infra-red degrees-of-freedom, from the microscopic, ultra-violet degrees-of-freedom. In a crude sense, the cut divides the Hilbert space into two factors $\BH_M\times\BH_E$, however, such a clean split of the Hilbert space is unlikely to be strictly true in realistic systems and so a more general definition of the cut will be needed.

\subsection{Decoherence and the polarization measurement}

When we include the environment in the our simple
model, the states of the measuring device $\ket{M_1}$ and $\ket{M_2}$ become entangled with orthogonal states of the environment, so that the final state now has has the form
\EQ{
\ket{\Psi(T)}=c_1\ket{M_1}\otimes\hat\Omega_1\ket{\Phi}\otimes\ket{E_1}+
c_2\ket{M_2}\otimes\hat\Omega_2\ket{\Phi}\otimes\ket{E_2}\ ,
}
where we have allowed for general measurements on the photon, e.g.~the noisy splitter \eqref{rrw}, via the POVM $\{\hat\Omega_1,\hat\Omega_2\}$. The reason why we can expect the states $\ket{E_i}$ to be orthogonal is that interactions are local in space, so the macroscopically distinct states $\ket{M_i}$ perturb a macroscopic number, order $N\sim10^{23}$ say, degrees-of-freedom of the environment in a distinct way. This leads to a suppression of the inner product $\bra{E_1}E_2\rangle$ by a factor $(1-\epsilon)^N$, for some small $\epsilon$. This is doubly exponentially small and it is safe to assume that the overlap vanishes (this argument is spelt out in \cite{Banks:2009gw}).

Now if we place the cut between $M$ and the photon-environment system, the local states are
\EQ{
\ket{M_1}\otimes\hat\Omega_1\ket{\Phi}\otimes\ket{E_1}\quad\text{and}\quad
\ket{M_2}\otimes\hat\Omega_2\ket{\Phi}\otimes\ket{E_2}\ .
}
The important point is that this is true even when the measurement operators $\hat\Omega_i$ are not projection operators. The reason is that the local states of the photon-environment subsystem, that is $\hat\Omega_i\ket{\Phi}\otimes\ket{E_i}$
are orthogonal because $\bra{E_1}E_2\rangle=0$ even when $\bra{\Phi}\hat\Omega_1^\dagger\hat\Omega_2\ket{\Phi}\neq0$.

So the environment solves the problem posed by generalized measurements in sect.~\ref{s4.2}. In addition, this conclusion is not affected if we decide to place the photon on the other side of the cut, showing in a modest way that the choice of cut is robust against moving a microscopic degree of freedom from one side to the other. This is something that will have to true more general if the cut is to be robust.

\subsection{Hyperion revisited}
  
We have argued that we cannot hide from \SCH cat states because they are generic in a chaotic system like Hyperion. The quantum phase space density is spread out and Ehrenfest's theorem will be violated. The quantum behaviour of the density is then very different from the corresponding classical density. Decoherence prevents this happening by essentially splitting the density into smaller decoherent pieces that don't ``talk" to each other and for which, individually, Ehrenfest's theorem is maintained. The implication is that decoherence ensures that the local states of the theory are appropriately quasi-classical even though the global state is spread out over phase space.

Before we tackle the technical issues that ensue in a quantum chaotic system, we must first clarify how to 
identify an appropriate cut in a realistic system. In particular, the identification of the cut with a strict factorization of the Hilbert space $\BH_M\times\BH_E$ is not general enough for our purposes.
The idea is to switch focus from the Hilbert space to the observables and to identify the cut via a set of preferred observables of a given resolution scale. This resolution scale is not to be thought as fixed but like the cut off in a quantum field theory can float as long as it is kept a sub-macroscopic scales. This is consistent because the predictions will be robust against changing the cut.

For Hyperion, decoherence is simply the process whereby the rotational state becomes entangled with the environment.
The latter represents all the microscopic external influences, like solar photons and interplanetary dust, along with the 
the moon's own internal microscopic constituents. The basic effect can be quantified by taking the expectation value of a non-local operator acting on the rotational system like
\EQ{
\hat A_\text{non-local}=e^{i\alpha}\ket{x,p}\bra{x',p'}+e^{-i\alpha}\ket{x',p'}\bra{x,p}\ ,
\label{ppp}
}
giving
\EQ{
\bra{\Psi}\hat A_\text{non-local}\ket{\Psi}=\RE\Big[e^{i\alpha}\bra{E_{x,p}}E_{x',p'}\rangle\Big]\quad\text{where}\quad\ket{E_{x,p}}=\bra{x,p}\Psi\rangle
} 
are states of the environment tagged by points in the rotational phase space. Decoherence is simply the phenomenon that the inner product of these environmental states falls off exponentially in phase space. In fact, detailed studies in toy models have shown that decoherence localizes on minimal $\sim\frac\hbar2$ area regions in phase space, i.e.~coherent states \cite{ZHP}, which means that the inner product above vanishes rapidly over spatial directions $\sim\ell_\text{coh}$, the coherence length, and $\wp_\text{coh}=\hbar/2\ell_\text{coh}$ in the momentum direction. The coherence length
depends on the detailed model of the interaction of Hyperion and the environment (be it external and/or internal).
In a simple model, where the decoherence is provided by collisions with interplanetary dust, a lower limit to the coherence length is the thermal de Broglie wavelength of Hyperion $\hbar/\sqrt{mk_BT}\sim10^{-34}\ \text{m}$ \cite{Joos:1984uk}.
(A more realistic estimate was made by Zurek \cite{Zurek2} for the planet Jupiter which when corrected for the mass of Hyperion gives $\sim\ 10^{-28}\ \text{m}$.) 
It is clear that whatever the exact coherence length, it is a scale  that is spectacularly small compared with any realistic macroscopic resolution scale. Whether realistic systems have sufficiently strong decoherence to localize on states with minimum area consistent with the uncertainty principle is not so important as the observation that the coherence scales in phase space are exceedingly small.

We now define a cut by specifying resolution scales $\ell_\text{cut}\times\wp_\text{cut}$ in phase space: 
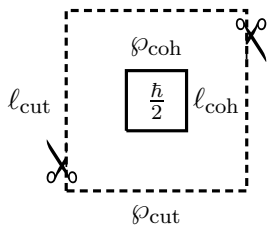
\begin{center}
\begin{tikzpicture}[scale=0.8,decoration={markings,mark=at position 0.6 with {\arrow{>}}}]
\draw[very thick] (0,0) -- (0,1) -- (1,1) -- (1,0) -- (0,0);
\draw[very thick,densely dashed] (-1,-1) -- (-1,2) -- (2,2) -- (2,-1) -- (-1,-1);
\node at (1.5,0.5) (a1) {$\ell_\text{coh}$};
\node at (0.5,1.4) (a2) {$\wp_\text{coh}$};
\node at (-1.6,0.5) (a3) {$\ell_\text{cut}$};
\node at (0.5,-1.4) (a4) {$\wp_\text{cut}$};
\node at (0.5,0.5) (a5) {$\frac\hbar2$};
\node[rotate=90] at (-1.05,-0.5) {\Huge\Cutright};
\node[rotate=-90] at (2.05,1.5) {\Huge\Cutright};
\end{tikzpicture}
\end{center}
As long as these resolution scales are much larger than the coherence scales, we can pick out the preferred set of local observables acting on Hyperion's rotational Hilbert space by partitioning phase space into compact disjoint subsets ${\EuScript A}_i$ of size $\ell_\text{cut}\times\wp_\text{cut}$ using the coherent states (with $\delta x=\ell_\text{coh}$) as
\EQ{
\hat\Pi_i=\int_{{\EuScript A}_i}\frac{dx\,dp}{2\pi\hbar}\,\ket{x,p}\bra{x,p}\ ,\qquad\sum_i\hat\Pi_i=1\ .
\label{rre}
}
The (over-)completeness of the coherent states \eqref{zza} means that the coarse-grained observables form a POVM.
All the other operators, like $\hat A_\text{non-local}$, for points $(x,p)$ and $(x',p')$ separated in phase space, fail as observables because 
\EQ{
\bra{\Psi}\hat A_\text{non-local}\ket{\Psi}\approx0\ .
}
The message here is that not all operators one can write down are really observables.

The observables \eqref{rre} are not quite a commuting set of projection operators, but as long as the resolution scales are kept well above the lower bounds set by the coherence scale they very nearly are. So the effects of decoherence and coarse graining leads to the {\it abelianization\/} (abelian=commuting) of the observables. 

Physics works by building models which allow a certain amount of flexibility in their definition. So my version of the standard model of particle physics might be slightly different from yours because we will probably cut off the high energy modes in different ways. (Note that the model is ill defined without such a cut off.) However, the predictions at low energy will be independent of these differences. The same will be true of the model of a macroscopic system in QM. You may partition up phase space in a different way from me; you may insist on exactly commuting observables whereas I will allow almost-commuting observables. It does not matter as long as the macroscopic predictions are robust against such model dependent differences.

The local states associated to cut are then given by the decomposition
\EQ{
\ket{\Psi}=\sum_i\ket{\Psi_i}\ ,\qquad \ket{\Psi_i}=\hat\Pi_i\ket{\Psi}\ .
}
The local states states will not be exactly orthogonal but this is in the spirit of building a model as we described above. Since the cut scale $\ell_\text{cut}\times\wp_\text{cut}$  is well below macroscopic scales, local states defined above are appropriately quasi classical, that is with support in localized regions of phase space.

\subsection{Is the cut subjective or objective?}

The cut obviously plays a key role and one might wonder whether it introduces an unacceptable  subjectivity. 
The first point to make is that the notion of a ``cut" separating out a set of relevant variables to a problem from irrelevant variables is a universal aspect of the way physics works.

In quantum field theory (QFT) this notion of a {\it separation of scales\/} has to be made very explicit. In this context the cut that separates the infra red modes from the ultra violet modes is known as the {\it Wilsonian cut off\/} \cite{Wilson:1983}. The cut off removes the ultra-violet modes at the expense of {\it renormalizing\/} the infra red modes, because the former are
not directly relevant for describing the low energy (long distance) phenomena that we can probe in an experiment. The Wilsonian cut off needs to be chosen beyond the resolution of our experiments but it is important that the scale of the cut can be changed without modifying the phenomena at longer distances (leading to the theory of the renormalization group). However, there is typically a limit to how small the cut can be taken for overall consistency. For example, the low energy theory of quantum chromodynamics is a field theory of pions. This theory only makes sense with a Wilsonian cut off up to a certain limit beyond which the description becomes inconsistent and one should switch to the more fundamental underlying theory. 

The lesson here is that the cut is to be placed somewhere arbitrarily below the resolution scale of the phenomena to be described but with some ultimate lower limit. This can provide a template for how to think of the Copenhagen cut. If we want to describe macroscopic behaviour then we should place the cut below macroscopic resolution scales. Beyond this, the cut is arbitrary but decoherence provides an ultimate lower limit to the resolution scales of measuring devices and so provides a natural short distance lower limit. 

So the position of the cut is really objective, if we want to describe macroscopic phenomena then this necessarily involves
specifying a certain resolution scale. This will always be much larger than any decoherence scale. The cut can then be placed arbitrarily between these scales, safe in the knowledge that any macroscopic predictions will be insensitive to the actual choice.

\section{Time dependence}\label{s6}

In this section we address the issue of the time dependence. The global state $\ket{\Psi(t)}$ evolves according to the Schr\"odinger equation but once the cut is specified the time dependence is inherited by the local states $\ket{\Psi_i(t)}$: 
\EQ{
\ket{\Psi(t)}=\sum_i\ket{\Psi_i(t)}\ ,\qquad p_i(t)=\bra{\Psi(t)}\Psi_i(t)\rangle\ .
}

Let us consider, once again, the photon polarization measurement. At $t=0$ the local state is uniquely $\ket{M_0}\otimes\ket{\Phi}$. In order to understand the time dependence of the local states, we need to specify the time dependence of the inner-products of the pointer states $\ket{M_i(t)}$. At $t=0$, they are equal equal to $\ket{M_0}$ but as time evolves they rapidly become orthogonal. We can expect that
\EQ{
F(t)=\big|\bra{M_i(t)}M_j(t)\rangle\big|\simeq\exp\big[-(t/\tau)^2\big]\ ,
\label{pxx}
}
where $\tau$ is a very short microscopic time scale.

It is then a simple matter to solve for the local states and probabilities as a function of $t$. The latter are
\EQ{
p_i(t)=\frac12\big(1\pm\sqrt{1-4|c_1|^2|c_2|^2(F(t)-1)^2}\big)\ .
}
If $|c_1|^2>|c_2|^2$, then at $t=0$ we have $p_1(0)=1$ and $p_2(0)=0$. At this stage, there is only one local state possible, equal, of course, to the global state $\ket{\Psi_1(0)}=\ket{M_0}\otimes\ket{\Phi}$. As $t$ evolves, up to a time $t\sim\tau$ the probabilites change and then settle down exponentially to their limiting values $|c_i|^2$ for $t>\tau$ and the local states become  $\ket{\Psi_1(t)}=\ket{M_1}\otimes\ket{h}$ and $\ket{\Psi(t)}=\ket{M_2}\otimes\ket{v}$.

\begin{figure}[ht]
\begin{center}
\begin{tikzpicture}[yscale=0.17,xscale=0.27]
\draw[thick] (0,0) -- (19,0) -- (19,19) -- (0,19) -- (0,0);
\node at (9.5,-2.2) (a1) {$t$};
\node at (-1.2,19) (a1) {$1$};
\node at (-1.2,0) (a1) {$0$};
\node at (7,13.2) (b1) {$p_1(t)$};
\node at (7,7.4) (b2) {$p_2(t)$};
\node at (15,13.2) (b1) {$\ket{M_1}\otimes\ket{h}$};
\node at (15,7.4) (b2) {$\ket{M_2}\otimes\ket{v}$};
\node at (6,22) (b4) {$\ket{M_0}\otimes\ket{\Phi}$};
\draw[->] (b4) -- (0.2,19.4);
\node[rotate=90] at (-2,10) (a1) {prob.};
\draw[-] (6.5,0) -- (6.5,0.5);
\node at (6.5,-1.7) (d1) {$\tau $}; 
\draw[very thick] plot[smooth] coordinates {(0, 19.)  (1, 18.2124)  (2, 16.4584)};
\draw[->,very thick]  (2, 16.4584) --  (2, 2.35062);
\draw[densely dashed] plot[smooth] coordinates {(2, 16.4584)  (3, 14.6642)  (4, 13.2187)  (5,12.2299)  (6, 11.6976)  (7,11.4838)  (8, 11.4189)  (9,11.4035)  (10, 11.4005)  (11, 11.4001)  (12, 11.4)  (13, 11.4)  (14,11.4)  (15, 11.4)  (16, 11.4)  (17, 11.4)  (18, 11.4)  (19, 11.4)};
\draw[densely dashed] plot[smooth] coordinates {(0, 0)  (1, 0.773204)  (2, 2.35062) };
\draw[very thick] plot[smooth] coordinates { (2, 2.35062)  (3, 3.65956) (4,4.51829)  (5, 5.12001)  (6, 5.48233)  (7, 5.6376)  (8, 5.68586)  (9,5.69741)  (10, 5.69961)  (11, 5.69995)  (12, 5.7)  (13, 5.7)  (14, 5.7)  (15, 5.7)  (16, 5.7)  (17, 5.7)  (18, 5.7)  (19, 5.7)};
\end{tikzpicture}
\end{center}
\caption{The time evolution of the probabilities during the polarization measurement showing a particular trajectory for the local state that includes a transition from $\ket{\Psi_1(t)}\to\ket{\Psi_2(t)}$. In this idealized situation the transitions occurs instantaneously (later we shall see in a realistic situation this the jump occurs over the temporal resolutions scale of the measuring device).}
\label{f6}
\end{figure}
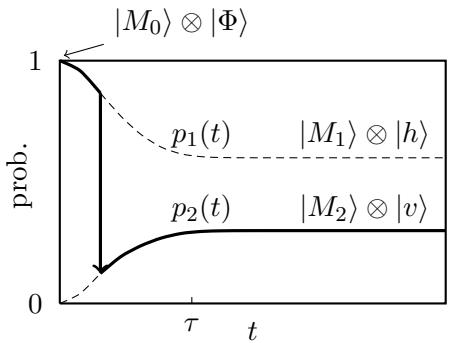

The question is, how could the system have ended up in the local state $\ket{M_2}\otimes\ket{v}$ which is {\it not\/} continuously connected to the initial state (when $|c_1|^2>|c_2|^2$)? The only way that this could have occurred is that 
in the region when the probabilities are changing there are transitions between the local states. An example of such a transition is shown in fig.~\ref{f6}.

\subsection{A quantum stochastic process}\label{s6.1}

The fact that there must be transitions between local states implies that there must be some underlying stochastic process. It might be thought that we do not really need the details of this {\it quantum stochastic process\/}, after all the transitions take place only in the very tiny window between $t=0$ and $t=\tau$ and all we need to know is the 
net result of the transitions, i.e.~the probabilities $|c_i|^2$, rather than the details. But this would be dodging an interesting phenomena and, in any case, there are other situations where the details {\it do\/} matter.

Our aim is to describe the process in terms of the instantaneous rates $T_{ji}$ for the transition $\ket{\Psi_i}\to\ket{\Psi_j}$.
In the following analysis, we only require that the local states are given by $\ket{\Psi_i}=\hat\Pi_i\ket{\Psi}$ for a POVM,
$\sum_i\hat\Pi_i=1$. In other words, it is not strictly necessary for the local states to be orthogonal. In addition, for simplicity, we will assume that the operators of the POVM are, themselves, time independent. The probabilities are given by
\EQ{
p_i(t)=\bra{\Psi(t)}\hat\Pi_i\ket{\Psi(t)}\ ,\qquad\sum_ip_i(t)=\bra{\Psi(t)}\sum_i\hat\Pi_i\ket{\Psi(t)}=1\ .
}

The transition rates must be consistent with the Schr\"odinger equation, from which we find
\EQ{
\frac{dp_i}{dt}&=\frac d{dt}\bra{\Psi(t)}\hat\Pi_i\ket{\Psi(t)}=
\frac i\hbar\bra{\Psi}\hat H\ket{\Psi_i}-\frac i\hbar\bra{\Psi_i}\hat H\ket{\Psi}\\
&=\frac i\hbar\sum_j\Big[\bra{\Psi_j}\hat H\ket{\Psi_i}-\bra{\Psi_i}\hat H\ket{\Psi_j}\Big]
=\frac2\hbar\sum_j\IM\,\bra{\Psi_i}\hat H\ket{\Psi_j}\ .
}
The transition rates should then be extracted by equating the right-hand side to $\sum_j\big(T_{ij}p_j-T_{ji}p_i\big)$, the net rate of transitions in and out of the $i^\text{th}$ local state. This, however, does not uniquely determine the rates. 
Fortunately there is a very simple and natural candidate for the rate that has sensible physical properties (first written down by Bell in a different context \cite{Bell:2004suqm}; further analyzed in the context of modal interpretations in \cite{BD10}; see also \cite{Sud1}):
\EQ{
T_{ji}=\frac2\hbar\,\text{max}\left[\frac{\IM\,\bra{\Psi_j}\hat H\ket{\Psi_i}}{\bra{\Psi}\hat\Pi_i\ket{\Psi}},0\right]\ .
\label{trs}
}
Note that the form above implies that if $T_{ij}>0$ then $T_{ji}=0$, and vice-versa, which will seen to be natural in the context of the decaying system analysed in sect.~\ref{s7.1}.

The most important properties of this process is (i) it is a Markov process (see below) defined by instantaneous transitions rates and (ii)  the instantaneous transition rates depend only the matrix elements of the Hamiltonian between the two local states. This is very natural since it ensures that the jump between the states is 
a spacetime local process, only possible for states that have non-vanishing matrix element with the Hamiltonian.  
\begin{quote}{\bf Markov process:} the simplest kind of stochastic process where the transition probabilities from the state at time $t$ only depend on the state at time $t$ and the states to which transitions are made but not on the state at earlier states. In a sense, the process has no memory, in that it does not care  how the system got to the current state. This allows the process to be defined by the instantaneous transition rates between states at a single instant of time.
\end{quote}

\subsection{Reduction of the wave function}\label{s6.2}

Recall that there are two notions of state, the global one and then the local one describing the state in the context of the cut. The global state plays an important role in determining the possible local states; however, in the measurement scenario we often don't need the full global state.
After the measurement, if the $i^\text{th}$ outcome occurs, this becomes imprinted on the environment as well as the macroscopic state of the measuring device. The question is whether there is any chance that in a future interaction, say another measurement, a local state imprinted with the $i^\text{th}$ result of the first measurement could make a transition to a local state imprinted with the $j^\text{th}$. According to the transition rates \eqref{trs} this would be impossible because the Hamiltonian could not have a non-vanishing matrix element between such macroscopically distinct states. The vanishing of certain transition rates implies the breaking of ergodicity of the quantum stochastic process, a phenomenon that we discuss more fully in sect.~\ref{s8.1}.

In terms of the quantum trajectories defined in sect.~\ref{s2.2}, at the $n^\text{th}$ measurement, transitions of the form
\EQ{
\ket{\Psi_{i_1,i_2,\ldots,i_n}}\longrightarrow \ket{\Psi_{j_1,j_2,\ldots,j_n}}\ ,
}
are only possible if $i_1=j_1$, $i_2=j_2$, etc., up to $i_{n-1}=j_{n-1}$: the past history of a trajectory cannot be changed.

For the stochastic dynamics of the local state, therefore, there is only need to keep track of a much smaller number of possible local states (the ones 
that are reachable) and, therefore, it makes sense to streamline the global state. In our polarization measurement, if the outcome is $h$, then, for the further evolution of the local state, we may replace the global state with $\ket{M_1}\otimes\ket{h}$. This is just the reduction of the wave function but now seen for what it really is, a booking keeping procedure useful for streamlining the calculation of the stochastic quantum trajectories of the local state.

Finally, the reduction of the wave function is only valid after a measurement has made and we should not fall into the trap of using it to infer the state before. So in our polarization measurement, if the result is $h$ then we cannot say ``the photon reflected off the beam splitter". The local state of the photon is still in the linear combination $c_1\ket{h}+c_2\ket{v}$ 
up to the point that the measurement occurs when the conditioned state changes to $\ket{h}$ (if the result $h$ is recorded). It is these retrodiction fallacies that make so-called {\it delayed choice\/} experiments appear to be weird.

\section{Bohr's Quantum Jumps}\label{s7}

Very early on in the development of quantum mechanics, Bohr described the interaction of light with atoms in terms of jumps between the quantized orbits of his atomic model \cite{BJump}. Later when QM was fully developed, it was understood how to calculate the rate at which an atom will decay often expressed as Fermi's Golden Rule but it is only comparatively recently that experimental techniques have evolved to the extent that we can observe single atoms undergoing spontaneous decay, making what looks, to all intents and purposes, like a quantum jump between energy levels in a completely random way \cite{NSD,SNBT,BHIW}. The same kind of phenomenon is observed when an atomic nucleus undergoes radioactive decay observed as a discrete process via a click in a Geiger counter.

What is fascinating is that these quantum jumps are {\it not\/} objective dynamical processes going on in the atom. Rather they result from the way the atom is monitored, i.e.~they describe the state of the atom conditioned on the local state of the measuring device. So just as we can change the context of the measurement of the photon by deciding to measure circular polarization rather than plane polarization, we can change the way we monitor the photon emitted by the atom so that instead of coming out all at once as a quantum jump it leaks out diffusively \cite{WG}. This, of course, is an example of Bohr's complementarity. 

\subsection{Spontaneous emission}\label{s7.1}

We will describe, first of all, how a conventional detection of the photon naturally gives rise to a quantum jump process when the Copenhagen cut is placed appropriately, i.e.~between the photon and the measuring device. The analysis in this section is taken from Sudbery \cite{Sud1}. 

Following the original analysis of Weisskopf and Wigner \cite{WW}, we will consider an atom in an excited state $\ket{\psi_1}$ decaying to the ground state $\ket{\psi_0}$ emitting a photon. The relevant states of the combined system are $\ket{\psi_1}\otimes\ket{0}$ and $\ket{\psi_0}\otimes\ket{\gamma(0)}$, where $\ket{0}$ is the electromagnetic vacuum and $\ket{\gamma(0)}$ is the 1-photon state just after emission from the atom. The decay occurs because the atom and radiation field are coupled by term in the Hamiltonian: 
\EQ{
\hat H_\text{int}=\lambda\Big\{\ket{\psi_0}\bra{\psi_1}\otimes\ket{\gamma(0)}\bra{0}+
\ket{\psi_1}\bra{\psi_0}\otimes\ket{0}\bra{\gamma(0)}\Big\}\ .
}

The solution of the Schr\"odinger equation can be written in the form
\EQ{
\ket{\Psi(t)}=f(t)e^{-iE_1t/\hbar}\ket{\psi_1}\otimes\ket{0}+
\frac{\lambda}{i\hbar}e^{-iE_0t/\hbar}\int_0^t dt'\,f(t')e^{-i\omega t'}\ket{\psi_0}\otimes \ket{\gamma(t-t')}\ ,
\label{nxx2}
}
where $\hbar\omega=E_1-E_0$ and $\ket{\gamma(t-t')}$ is the state of the 
radiation field describing a photon produced at $t'$ and propagating away for a time $t-t'$.
The function $f(t)$ satisfies the first order equation integral equation
\EQ{
\frac{df(t)}{dt}=-\frac{\lambda^2}{\hbar^2}\int_0^tdt'\,q(t-t')f(t')\ ,\qquad q(t)=\bra{\gamma(0)}\gamma(t)\rangle e^{i\omega t}\ .
\label{ddu}
}
The first term of \eqref{nxx2} is the excited state at time $t$ while the second term corresponds to a decay at time $t'$ to $\ket{\psi_0}\otimes\ket{\gamma(0)}$ and then the propagation of the photon state for time $t-t'$.

In order to make analytic progress we can make the {\it rapid dispersal approximation\/} \cite{Sud}. 
The key to that lies in the fact that
the photon rapidly leaves the vicinity of the atom over a time scale that is much smaller than the scale over which $f(t)$ varies. This implies that the time evolved photon state $\ket{\gamma(t)}$ becomes rapidly orthogonal to $\ket{\gamma(0)}$. We will suppose that this occurs on a time scale that is much shorter than any other scale in the problem, in which case we can assume
\EQ{
q(t)=\tau\delta(t)\ ,
\label{op1}
}
where $\tau$ is some time scale characteristic of the rapid dispersal process. In that case \eqref{ddu}, becomes
\EQ{
\frac{df(t)}{dt}=-\gamma f(t)\ ,\qquad \gamma=
\frac{\lambda^2\tau}{2\hbar^2}\ .
\label{gyq}
}
Hence $f(t)=e^{-\gamma t}$ and so the inverse lifetime of the excited state is 
\EQ{
2\gamma=\frac\tau{\hbar^2}\big|\bra{\psi_0}\otimes\bra{\gamma(0)}\,\hat H_\text{int}\,\ket{\psi_1}\otimes\ket{0}\big|^2\ ,
}
which is a version of Fermi's Golden Rule.

It is tempting to suppose that we could introduce a cut between the atom and the photon to define local states. However, this leads to a phenomenological disaster, local states that are not localized in the decay time $t'$. This would not produce  the observed quantum jumps \cite{Hollowood:2013cbr}. 

In retrospect, of course, we shouldn't expect to get a correct phenomenological description without introducing a macroscopic measuring device to detect the radiation field. The simplest kind of detector simply registers the arrival time of the photon with some realistic finite temporal resolution $\eta$, which we assume is much smaller than the atomic lifetime $\eta\gamma\ll 1$. Therefore, $M$ has pointer states $\ket{M_j}$ corresponding to a photon  arriving in a small window of time $(j-1)\eta<t<j\eta$. 

Putting the cut between $M$ and the atom plus radiation field, gives local states
\EQ{
\ket{\Psi_j(t)}=\ket{M_j(t)}\otimes
\frac{\lambda}{i\hbar}e^{-iE_0t/\hbar}\int_{(j-1)\eta}^{j\eta} dt'\,e^{-\gamma t'-i\omega t'}\ket{\psi_0}\otimes\ket{\gamma(t-t')}\ .
}
In the above, we have not explicitly included the environment. It ensures that the local states of the measuring device are quasi classical. We have also not indicated how the measuring device acts on the photon state (presumably the photon is absorbed and the radiation state returns to the vacuum). In addition, there is the local state 
\EQ{
\ket{\Psi_0(t)}=\ket{M_0(t)}\otimes e^{-\gamma t-iE_1t/\hbar}\ket{\psi_1}\otimes\ket{0}\ ,
}
which corresponds to no decay up to time $t$.

The probability of the $j^\text{th}$ outcome is found to be
\EQ{
p_0=\bra{\Psi_0}{\Psi_0}\rangle=e^{-2\gamma t}\ ,\qquad p_j=\bra{\Psi_j}\Psi_j\rangle=2\gamma\eta e^{-2\gamma j\eta}\ .
}
Note that $\sum_{j=0}^{n}p_j=1$ when $\eta\gamma\ll1$ and $n=t/\eta$.

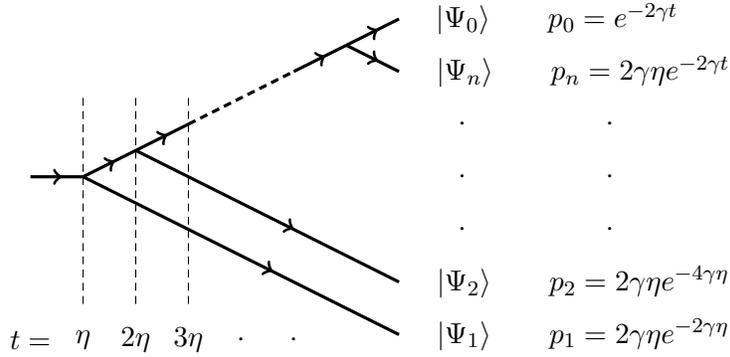
\begin{figure}[ht]
\begin{center}
\begin{tikzpicture}[scale=0.7,decoration={markings,mark=at position 0.6 with {\arrow{>}}}]
\draw[very thick,postaction={decorate}] (0,0.5) -- (1,0.5);
\draw[very thick,densely dashed] (3,1.5) -- (5,2.5);
\draw[very thick,postaction={decorate}] (1,0.5) -- (2,1);
\draw[very thick,postaction={decorate}] (2,1) -- (3,1.5);
\draw[very thick,postaction={decorate}] (5,2.5) -- (6,3);
\draw[very thick,postaction={decorate}] (6,3) -- (7,3.5);
\draw[very thick,postaction={decorate}] (1,0.5) -- (7,-2.5);
\draw[very thick,postaction={decorate}] (2,1) -- (7,-1.5);
\draw[very thick,postaction={decorate}] (6,3) -- (7,2.5);
\node at (10,3.5) (a1) {$\ket{\Psi_0}\qquad p_0=e^{-2\gamma t}$};
\node at (10.5,2.5) (a1) {$\ket{\Psi_n}\qquad p_n=2\gamma\eta e^{-2\gamma t}$};
\node at (10.5,-1.5) (a1) {$\ket{\Psi_2}\qquad p_2=2\gamma\eta e^{-4\gamma\eta}$};
\node at (10.5,-2.5) (a1) {$\ket{\Psi_1}\qquad p_1=2\gamma\eta e^{-2\gamma\eta}$};
\node at (8.2,1.5) (b1) {$\cdot$};
\node at (8.2,0.5) (b1) {$\cdot$};
\node at (8.2,-0.5) (b1) {$\cdot$};
\node at (11,1.5) (b1) {$\cdot$};
\node at (11,0.5) (b1) {$\cdot$};
\node at (11,-0.5) (b1) {$\cdot$};
\draw[densely dashed] (1,2) -- (1,-2);
\draw[densely dashed] (2,2) -- (2,-2);
\draw[densely dashed] (3,2) -- (3,-2);
\node at (1,-2.6) (c1) {$\eta$};
\node at (2,-2.6) (c2) {$2\eta$};
\node at (3,-2.6) (c1) {$3\eta$};
\node at (4,-2.6) (c2) {$\cdot$};
\node at (5,-2.6) (c2) {$\cdot$};
\node at (0,-2.6) (c1) {$t=$};
\end{tikzpicture}
\end{center}
\caption{The stochastic process for the decaying system with a finite temporal resolution $\eta$ up to time $t=n\eta$. The local state $\ket{\Psi_j}$ indicates that the atom decayed in the time window $(j-1)\eta<t<j\eta$.}
\label{f7}
\end{figure}

The quantum stochastic process, illustrated in fig.~\ref{f7}, in this case is simple. In each time window, the local state $\ket{\Psi_0}$, the excited state, can make a transition to local state $\ket{\Psi_j}$ with a rate $2\gamma$. Since the excited state 
only survives to time $j\eta$ with a probability $e^{-2\gamma j\eta}$, the probability that the transition is made to $\ket{\Psi_j}$ is $p_j$ written above.

The basic jump of the atom corresponds to the stochastic evolution over a single time step with measurement operators
\EQ{
\hat\Omega_1=1-\big(i\hat H+\gamma a^\dagger a\big)\eta\ ,\qquad\hat\Omega_0=\sqrt{2\gamma\eta}\,a\ ,
}
to order $\eta$, where we have defined $a=\ket{\psi_0}\bra{\psi_1}$. The trajectories in fig.~\ref{f7} with probability $p_j$ correspond to the conditioned (local) states
\EQ{
\big(\hat\Omega_1\big)^n\ket{\psi_1}\quad (j=0)\ ,\qquad\big(\hat\Omega_1\big)^{n-j}\hat\Omega_0\big(\hat\Omega_1\big)^{j-1}\ket{\psi_1}\quad(j=1,2,\ldots,n)\ .
}
Note that the only possible trajectories are ones with at most one $\hat\Omega_1$ because
$\hat\Omega_1(\hat\Omega_0)^j\hat\Omega_1=0$. In contrast to the description in sect.~\ref{s2.2}, here the measurements and unitary evolution are mixed together in each time step on account of the finite resolution. 

\subsection{Different measurements}

\begin{figure}[ht]
\begin{center}
\begin{tikzpicture}[scale=4]
\draw[thick] (0,0) -- (1.6,0) -- (1.6,1) -- (0,1) -- (0,0);
\draw[thick] (0,0.5) -- (0.04,0.5);
\node at (-0.1,0) {$0$};
\node at (-0.1,0.5) {$0.5$};
\node at (-0.1,1) {$1$};
\node[rotate=90] at (-0.3,0.5) {overlap};
\node at (0.8,-0.07) {$t$};
\begin{scope}[yshift=1cm]
\draw[thick] plot coordinates {(0.016,-0) (0.032,-0.0196) (0.048,-0.0392) (0.064,-0.0589) (0.08,-0.0786) (0.096,-0.0983) (0.112,-0.118) (0.128,-0.0659) (0.144,-0.273) (0.16,-0.257) (0.176,-0.241) (0.192,-0.224) (0.208,-0.207) (0.224,-0.189) (0.24,-0.172) (0.256,-0.154) (0.272,-0.136) (0.288,-0.117) (0.304,-0.099) (0.32,-0.0803) (0.336,-0.0617) (0.352,-0.0431) (0.368,-0.0256) (0.384,-0.0969) (0.4,-0.389) (0.416,-0.375) (0.432,-0.361) (0.448,-0.347) (0.464,-0.332) (0.48,-0.317) (0.496,-0.302) (0.512,-0.287) (0.528,-0.271) (0.544,-0.255) (0.56,-0.239) (0.576,-0.222) (0.592,-0.206) (0.608,-0.189) (0.624,-0.131) (0.64,-0.504) (0.656,-0.493) (0.672,-0.483) (0.688,-0.472) (0.704,-0.46) (0.72,-0.449) (0.736,-0.437) (0.752,-0.425) (0.768,-0.412) (0.784,-0.399) (0.8,-0.386) (0.816,-0.373) (0.832,-0.36) (0.848,-0.346) (0.864,-0.332) (0.88,-0.318) (0.896,-0.303) (0.912,-0.289) (0.928,-0.149) (0.944,-0.563) (0.96,-0.555) (0.976,-0.546) (0.992,-0.537) (1.01,-0.527) (1.02,-0.518) (1.04,-0.19) (1.06,-0.699) (1.07,-0.217) (1.09,-0.799) (1.1,-0.796) (1.12,-0.794) (1.14,-0.792) (1.15,-0.789) (1.17,-0.786) (1.18,-0.784) (1.2,-0.225) (1.22,-0.847) (1.23,-0.845) (1.25,-0.844) (1.26,-0.842) (1.28,-0.23) (1.3,-0.882) (1.31,-0.881) (1.33,-0.88) (1.34,-0.879) (1.36,-0.878) (1.38,-0.878) (1.39,-0.877) (1.41,-0.876) (1.42,-0.875) (1.44,-0.23) (1.46,-0.902) (1.47,-0.901) (1.49,-0.901) (1.5,-0.901) (1.52,-0.9) (1.54,-0.9) (1.55,-0.899) (1.57,-0.899) (1.58,-0.898) (1.6,-0.898)};
\draw[thick] plot coordinates {(0.016,-0) (0.032,-0.000652) (0.048,-0.00131) (0.064,-0.00196) (0.08,-0.00262) (0.096,-0.00329) (0.112,-0.00395) (0.128,-0.00462) (0.144,-0.00529) (0.16,-0.00597) (0.176,-0.00664) (0.192,-0.00732) (0.208,-0.00801) (0.224,-0.00869) (0.24,-0.00938) (0.256,-0.0101) (0.272,-0.0108) (0.288,-0.0115) (0.304,-0.0122) (0.32,-0.0129) (0.336,-0.0136) (0.352,-0.0143) (0.368,-0.015) (0.384,-0.0157) (0.4,-0.0164) (0.416,-0.0171) (0.432,-0.0179) (0.448,-0.0186) (0.464,-0.0193) (0.48,-0.02) (0.496,-0.0208) (0.512,-0.0215) (0.528,-0.0222) (0.544,-0.023) (0.56,-0.0237) (0.576,-0.0245) (0.592,-0.0252) (0.608,-0.026) (0.624,-0.0267) (0.64,-0.0275) (0.656,-0.0283) (0.672,-0.029) (0.688,-0.0298) (0.704,-0.0306) (0.72,-0.0313) (0.736,-0.0321) (0.752,-0.0329) (0.768,-0.0337) (0.784,-0.0345) (0.8,-0.0352) (0.816,-0.036) (0.832,-0.0368) (0.848,-0.0376) (0.864,-0.0384) (0.88,-0.0392) (0.896,-0.04) (0.912,-0.0408) (0.928,-0.0416) (0.944,-0.0425) (0.96,-0.0433) (0.976,-0.0441) (0.992,-0.0449) (1.01,-0.0918) (1.02,-0.997) (1.04,-0.997) (1.06,-0.997) (1.07,-0.997) (1.09,-0.997) (1.1,-0.997) (1.12,-0.997) (1.14,-0.997) (1.15,-0.997) (1.17,-0.997) (1.18,-0.997) (1.2,-0.997) (1.22,-0.997) (1.23,-0.997) (1.25,-0.997) (1.26,-0.997) (1.28,-0.997) (1.3,-0.997) (1.31,-0.997) (1.33,-0.997) (1.34,-0.997) (1.36,-0.997) (1.38,-0.997) (1.39,-0.997) (1.41,-0.998) (1.42,-0.998) (1.44,-0.998) (1.46,-0.998) (1.47,-0.998) (1.49,-0.998) (1.5,-0.998) (1.52,-0.998) (1.54,-0.998) (1.55,-0.998) (1.57,-0.998) (1.58,-0.998) (1.6,-0.998)};
\draw[thick] plot coordinates {(0.016,-0) (0.032,-0.0922) (0.048,-0.0109) (0.064,-0.104) (0.08,-0.0218) (0.096,-0.0496) (0.112,-0.121) (0.128,-0.191) (0.144,-0.257) (0.16,-0.318) (0.176,-0.372) (0.192,-0.419) (0.208,-0.46) (0.224,-0.495) (0.24,-0.459) (0.256,-0.492) (0.272,-0.522) (0.288,-0.548) (0.304,-0.569) (0.32,-0.588) (0.336,-0.57) (0.352,-0.586) (0.368,-0.602) (0.384,-0.615) (0.4,-0.626) (0.416,-0.621) (0.432,-0.626) (0.448,-0.635) (0.464,-0.643) (0.48,-0.65) (0.496,-0.656) (0.512,-0.662) (0.528,-0.673) (0.544,-0.661) (0.56,-0.673) (0.576,-0.661) (0.592,-0.666) (0.608,-0.68) (0.624,-0.673) (0.64,-0.665) (0.656,-0.656) (0.672,-0.661) (0.688,-0.665) (0.704,-0.669) (0.72,-0.673) (0.736,-0.676) (0.752,-0.679) (0.768,-0.703) (0.784,-0.679) (0.8,-0.681) (0.816,-0.683) (0.832,-0.685) (0.848,-0.686) (0.864,-0.688) (0.88,-0.721) (0.896,-0.688) (0.912,-0.689) (0.928,-0.69) (0.944,-0.727) (0.96,-0.69) (0.976,-0.691) (0.992,-0.729) (1.01,-0.727) (1.02,-0.69) (1.04,-0.691) (1.06,-0.692) (1.07,-0.693) (1.09,-0.694) (1.1,-0.736) (1.12,-0.694) (1.14,-0.694) (1.15,-0.695) (1.17,-0.696) (1.18,-0.696) (1.2,-0.696) (1.22,-0.697) (1.23,-0.697) (1.25,-0.698) (1.26,-0.698) (1.28,-0.698) (1.3,-0.698) (1.31,-0.75) (1.33,-0.698) (1.34,-0.699) (1.36,-0.699) (1.38,-0.699) (1.39,-0.699) (1.41,-0.754) (1.42,-0.699) (1.44,-0.754) (1.46,-0.699) (1.47,-0.754) (1.49,-0.699) (1.5,-0.699) (1.52,-0.755) (1.54,-0.699) (1.55,-0.699) (1.57,-0.699) (1.58,-0.7) (1.6,-0.757)};
\end{scope}
\node at (1.9,0.5) {\phantom{.}};
\end{tikzpicture}
\end{center}
\caption{Three simulated quantum trajectories for the conditioned state of the atom $\ket{\psi(t)}$ starting in the excited state $\ket{\psi_1}$ as measured by the overlap $\big|\bra{\psi_1}\psi(t)\rangle\big|$ for different values of $\beta$. For small $\beta$ the trajectory has an obvious jump which we interpret as one of Bohr's quantum jumps, while as $\beta$ increases the trajectories consist of a series of smaller jumps and eventually become smooth and the idea of a photon being emitted randomly becomes untenable.}
\label{f9}
\end{figure}
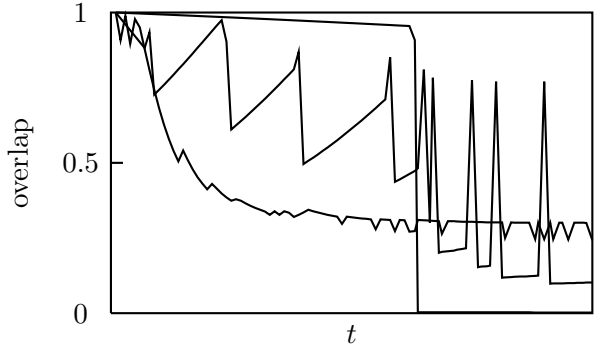

We should not fall into the trap of thinking that this stochastic evolution is an objective dynamical process going on inside the atom. Rather it is the dynamics of the state {\it conditioned\/} on the measurement of the arrival time of the photon. To ram this point home, we can change the way the radiation field is measured, for instance by replacing direct photon detection with {\it homodyne\/} detection (e.g.~\cite{WG}). This is achieved by interfering the output field with an external  coherent field using a suitable beam splitter. The new apparatus no longer measures the arrival time of a photon and the conditioned evolution of the atom is now described by modified measurement operators
\EQ{
\hat\Omega_1=1-\big(i\hat H+2\sqrt\gamma\beta^*a+\gamma a^\dagger a+|\beta|^2\big)\eta\ ,\qquad\hat\Omega_0=\sqrt{2\eta}\,\big(\sqrt\gamma a+\beta\big)\ ,
}
to order $\eta$, where $\beta$ describes the strength and phase of the coherent field. 
With this measurement context there is much richer set of possible trajectories. In particular, in the limit where the coherent field becomes very strong, i.e.~large $\beta$, both $\hat\Omega_0$ and $\hat\Omega_1$ become close to the identity
(up to an overall factor). So in this limit, the quantum trajectories are no longer jumpy but become smooth and diffusive. Some simulated trajectories for different values of $\beta$ are shown in fig.~\ref{f9}. However, we should emphasize that whatever measurement scheme is chosen, the underlying stochastic dynamics is the quantum stochastic process of sect.~\ref{s6.1}.

So Bohr's quantum jumps describe how the state of the atom evolves conditioned on the result of the measurement of the arrival time of the photon. Without such a measurement there is no jump because there is nothing to condition the state. On the other hand, it is not correct to say that the measuring device actually caused the jump, after all that happens randomly. The correct way to think about this is that the measuring device provides a context in which jumps, rather than some other form of evolution, can happen, providing a perfect illustration of Bohr's notion of complementarity.

\section{Classical Emergence}\label{s8}

It is ironic, perhaps, that the key to classical behaviour lies in the entanglement, the most quantum of phenomena, between a macroscopic system and its much larger environment. However, can we account for the different kinds of classical behaviour? On the one hand, we have classical mechanics, which involves systems following trajectories in phase space while, on the other, we have thermal behaviour that involves statistical ensembles and densities distributed over phase space. Finally we have a mixture of the two that we see during a quantum measurement.

\subsection{A useful analogue: the Ising model}\label{s8.1}

The Ising model of a magnet provides a very useful analogue for the behaviour of macroscopic systems in Copenhagen QM. The state of the model is defined at the microscopic level by spins $s_\mu\in\{\pm1\}$ (either up or down) at the sites of a lattice labelled by $\mu$. The set of spins defines the microstate, the analogue of the local state of a macroscopic system in QM.

The dynamics of the spins is defined by a stochastic process which models the interactions of the spins with a heat bath. There is no unique choice of process, the resulting behaviour is insensitive to this freedom.
For instance, one choice is Glauber's dynamics \cite{Glauber} which is defined in terms of the probability for the $\mu^\text{th}$ spin to flip in a given short interval of time:
\EQ{
p(s_\mu\to-s_\mu)=\Big[1+\exp\big(s_\mu\sum_{\nu(\mu)}s_\nu/T\big)\Big]^{-1}\ ,
}
where $T$ is the temperature and the sum is over the nearest neighbours to the $\mu^\text{th}$ spin. Glauber dynamics is the analogue of the quantum stochastic process. Both are Markov processes.

The Ising model is very simple, but illustrates a characteristic feature that carries over into the QM context. The probability of a spin flip favours a given spin pointing in the same direction as the average spin of its neighbours, an effect that is magnified at lower temperatures.
When the temperature is high enough, the dynamics is {\it ergodic\/} and over time cycles though the complete set of microscopic states. In particular, for long times the system has no net magnetization: a given spin spends roughly equal amounts of time pointing up or down. When the temperature drops below a critical temperature, the Curie temperature, {\it ergodicity\/} is broken. If the average spin is positive, then over time the system only has a very small probability to jump into an average negative spin state, and vice-versa. The breaking of ergodicity implies that the dynamics only explores only half the available states over time. 

This behaviour provides a simple template for the behaviour of macroscopic systems in QM. In particular, let us return to the \SCH cat state \eqref{scs}, but now including implicitly the environment. Because of decoherence, the Copenhagen cut is associated to a quasi-classical POVM $\{\hat\Pi_i\}$ (that is a set of almost commuting projection operators associated to coarse-grained regions in the cat's phase space.) The local states are then given by
\EQ{
\ket{\Psi_i}=\hat\Pi_i\big(c_1\ket{\LCATH}+c_2\ket{\DCATH}\big)\ .
\label{scs2}
}
If we denote the set of phase space regions as $S$, then
because of the locality of the observables in phase space, only disjoint subsets of the complete set $S_a\subset S$, $a=1,2$, are non-vanishing with
\EQ{
\ket{\Psi_i}=\begin{cases}c_1\hat\Pi_i\ket{\LCATH} \quad\qquad&i\in S_1\ ,\\[7pt] c_2\hat\Pi_i\ket{\DCATH} &i\in S_2\ .\end{cases}
}

These subsets are the analogues of the subsets of net up and net down spin states of the Ising model below the Curie temperature. Like the Ising model, ergodicity is broken. There is zero chance of a local state of a dead cat making a transition to a live cat state, and vice-versa:
\EQ{
\bra{\DCATH}\hat\Pi_i\hat H\hat\Pi_j\ket{\LCATH}=0\ .
}
On the other hand, there will be transitions between the multitude of local states of a live (or a dead) cat.

A macroscopic object like a cat, corresponds to set of local states whose quantum stochastic process describes the classical dynamics of the cat (if alive) but also must account for its internal thermal properties.
The former kind of behaviour corresponds to directions in phase space where the quantum stochastic process breaks ergodicity strongly. In such a situation a sequence of local states is uniquely defined by the initial state, so the dynamics is effectively deterministic. In the next section we turn to the question of thermal behaviour.

\subsection{Thermodynamics}\label{s8.2}

QM has shown that it can provide a consistent microscopic underpinning of classical statistical mechanics and thermodynamics (some selected references are \cite{PopescuShortWinter:2006efsm,GoldsteinLebowitzTumulkaZanghi:2006ct,GMM,Reimann1,Srednicki2}). What these developments show is that thermal behaviour is captured by the overall state vector, what we have called the global state. There seems to be no room for the local states in this approach. Is this a problem for the very notation of the local states?

Suppose that the global state $\ket{\Psi}$ is any state which lies in the micro-canonical subspace $\BH_\text{mc}$, the subspace of states with energy lying in a narrow band between $E$ and $E+\Delta E$. If we expand the state in terms of energy eigenstates, 
\EQ{
\ket{\Psi(t)}=\sum_kc_ke^{-itE_k/\hbar}\ket{\psi_k}\ ,\qquad E_k\subset[E,E+\Delta E]\ ,
}
then it is clear that the state itself can never equilibrate, i.e.~become time independent. 
The question of whether the system equilibrates is not determined by the time dependence of the state $\ket{\Psi(t)}$ itself, but whether expectation values of the macroscopic observables $\bra{\Psi(t)}\hat\Pi_i\ket{\Psi(t)}$ 
become constant for times in the long run (very occasional excursions away from equilibrium are to be expected). 

There is a more refined question of whether the system thermalizes, that is whether 
the expectation values $\bra{\Psi(t)}\hat\Pi_i\ket{\Psi(t)}$ equilibrate to their micro-canonical values. 
The assumption is that each $\hat\Pi_i$ acts as a projection operator on the micro-canonical subspace $\BH_\text{mc}$ of the full Hilbert space. The micro-canonical value for the expectation value of $\hat\Pi_i$ is then simply the fractional size of the subspace $\hat\Pi_i\BH_\text{mc}$:
\EQ{
\bra{\Psi(t)}\hat\Pi_i\ket{\Psi(t)}=\frac{d_i}{d_\text{mc}}\ ,
\label{nxx}
}
where $d_i$ is the dimension of the subspace $\hat\Pi_i{\cal H}_\text{mc}$ and $d_\text{mc}$ is the dimension of ${\cal H}_\text{mc}$ and $\sum_id_i=d_\text{mc}$. 
 
This is exactly what Von Neumann proved in 1929 in his {\it quantum ergodic theorem\/} \cite{vN} for most states $\ket{\Psi(t)}\in\BH_\text{mc}$ and most times in the long run (see also \cite{GLMTZ}). It also follows from the more powerful {\it eigenstate thermalization hypothesis\/} \cite{Srednicki2}.
 
The key thing here, is that it is expectation values with respect to the {\it global\/} state that has thermal properties. 
So this is a situation where the expectation values with respect to one of the {\it local\/} states would give the wrong answer. The reason is that the local states are the simultaneous eigenstates of the set $\{\hat\Pi_i\}$ with eigenvalue 1 for one element of the set and 0 for all the others. If this eigenstate were the $j^\text{th}$ then $\bra{\Psi_j}\hat\Pi_i\ket{\Psi_j}=\delta_{ij}$ and not equal to the micro-canonical value of $d_i/d_\text{mc}$.

\begin{figure}[ht]
\begin{center}
\begin{tikzpicture}[scale=0.6]
\node at (-10.2,1.7) (a1) {$j(t)=$};
\node at (-6.7,3.2) (a2) {$j_2\qquad t<t_1$};
\node at (-6,2.2) (a2) {$j_4\qquad t_1<t<t_2$};
\node at (-6,1.2) (a2) {$j_1\qquad t_2<t<t_3$};
\node at (-6,0.2) (a2) {$\cdots$};
\draw[decoration={brace,amplitude=0.5em},decorate,very thick] (-8.7,0) -- (-8.7,3.5);
\draw[densely dashed] (0,0) -- (8,0);
\draw[densely dashed] (0,1) -- (8,1);
\draw[densely dashed] (0,2) -- (8,2);
\draw[densely dashed] (0,3) -- (8,3);
\node at (-1,0) (a1) {$\ket{\Psi_{j_1}}$};
\node at (-1,1) (a1) {$\ket{\Psi_{j_2}}$};
\node at (-1,2) (a1) {$\ket{\Psi_{j_3}}$};
\node at (-1,3) (a1) {$\ket{\Psi_{j_4}}$};
\node at (2,-0.8) (h1) {$t_1$};
\node at (4,-0.8) (h1) {$t_2$};
\node at (5.5,-0.8) (h1) {$t_3$};
\draw[very thick] (0,1) -- (2,1);
\draw[decorate,decoration={snake,amplitude=0.05cm},very thick,->] (2,1) -- (2,3);
\draw[very thick] (2,3) -- (4,3);
\draw[decorate,decoration={snake,amplitude=0.05cm},very thick,->] (4,3) -- (4,0);
\draw[very thick] (4,0) -- (5.5,0);
\draw[decorate,decoration={snake,amplitude=0.05cm},very thick,->] (5.5,0) -- (5.5,2);
\draw[very thick] (5.5,2) -- (8,2);
\end{tikzpicture}
\end{center}
\caption{The quantum stochastic process leads to quantum trajectories $\ket{\Psi_{j(t)}}$ where the label $j(t)$ changes discontinuously at random times according to the rates $T_{ij}$.}
\label{f8}
\end{figure}
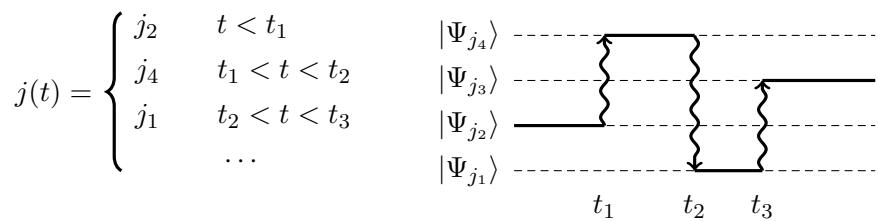

This would be a problem if we were not missing the fact that the local state is not fixed but, rather, undergoes a stochastic evolution. So over time, there will be a quantum trajectory. Let us describe a trajectory as $\ket{\Psi_{j(t)}}$, where $j(t)$ is a discontinuous function that describes the random jumps: see fig.~\ref{f8}. There is a subtle point here: unlike the quantum trajectories we described in sect.~\ref{s2.2}, the local states do not have any memory of earlier states in the trajectory because here the system is interacting with an thermal bath rather than a measuring device which keeps track of the trajectory.
In particular, when the system has thermalized, \eqref{nxx} implies that the probabilities associated to the local states become constant:
\EQ{
p_i=\bra{\Psi(t)}\hat\Pi_i\ket{\Psi(t)}=\frac{d_i}{d_\text{mc}}\ .
}

What really matters then is the long time average of the trajectory of expectation values, i.e.~for some suitable long time $T$
\EQ{
\frac1T\int_0^T dt\,\bra{\Psi_{j(t)}}\hat\Pi_i\ket{\Psi_{j(t)}}=\text{fraction of time spent in }i^\text{th}\text{ state}\ .
\label{uus}
}
So the question is whether the fraction of time spent in the $i^\text{th}$ state over a long time is equal to the micro-canonical average $d_i/d_\text{mc}$? 

This is precisely the question of whether the quantum stochastic process is {\it ergodic\/}. Stochastic processes of the type we have (i.e.~a Markov process) are ergodic if over time the process cycles through all its states. This requires that any state can reach any other state in a finite number of steps. Equivalently, an ergodic process is one which forgets its initial state. Under these circumstances, it follows that the long time average \eqref{uus} will equal to single time probability $p_i$, i.e.~\eqref{uus} will indeed equal its micro-canonical value.

So is the quantum stochastic process ergodic? Recall from \eqref{trs} that the transition rates between the local states involve the matrix element of the Hamiltonian $\bra{\Psi_i}\hat H\ket{\Psi_i}$. On general grounds, in an interacting, i.e.~chaotic, system it is reasonable to expect that the mutual eigenstates of the set $\{\hat\Pi_i\}$ and the Hamiltonian are related in a completely random way. The intuition here is that the stationary states have a quantum phase space density that is spread out ergodically in phase space, whereas eigenstates of $\hat\Pi_i$ are localized in phase space. This means that the transition rates are essentially random. In these circumstances, the quantum stochastic process {\it is\/} indeed ergodic.

\section{Discussion}\label{s9}

The Copenhagen interpretation that we learn in our undergraduate QM courses may not do justice to the subtleties of the original QM of the Copenhagen pioneers. However, it is perfectly good set of ``rules-of-thumb" for dealing with macroscopic objects in quantum mechanics. Essentially this rule says that when a macroscopic object is involved we should treat its quantum phase space density as a classical density at the coarse-grained level compared with the coherence scale. So macroscopic superpositions of a dead and live cat correspond to ``or" and not ``and". These rules-of-thumb work remarkably  well and are essential for understanding the modern theory of measurement and the notion of the conditioned state and quantum trajectories. 

However, we have argued that the authentic 1920's QM, rather than the later 1950's re-working, when mixed with the insights of decoherence, does provide a consistent microscopic understanding from which the Copenhagen interpretation rules-of-thumb emerge in the macroscopic realm. The approach brings the stochastic nature of the real world to the forefront rather than treating it as a secondary concern.


\end{document}